\documentclass[11pt,a4paper]{article}
\usepackage[utf8]{inputenc}
\usepackage[T1]{fontenc}
\usepackage{lmodern}
\usepackage[margin=2.5cm]{geometry}
\usepackage{amsmath,amssymb}
\usepackage{graphicx}
\usepackage{booktabs,longtable,multirow,array}
\newcolumntype{L}[1]{>{\raggedright\arraybackslash}p{#1}}
\usepackage{pdflscape}
\usepackage[table]{xcolor}
\usepackage{setspace}
\usepackage[font=small,labelfont=bf]{caption}
\usepackage[hidelinks]{hyperref}
\usepackage{xurl}

\graphicspath{{figs/}}
\hypersetup{
  pdftitle={Seismic study of soda straws exposed to nearby blasting vibrations},
  pdfauthor={P. Bottelin, L. Baillet, A. Mathy, L. Garnier, H. Cadet, O. Brenguier},
  pdfsubject={Preprint},
  pdfkeywords={Blasting works, Rock slope cutting, Speleothems, Numerical modeling, Dynamic amplification, Vibration thresholds}
}

\newcommand{\sut}{\sigma_{u,t}}
\newcommand{\sutmax}{\sigma_{u,t,\max}}
\newcommand{\sutd}{\sigma_{u,t,d}}
\newcommand{\GPC}{GP3C}
\newcommand{\GPCten}{GP3C$_{10}$}

\begin{document}

\begin{center}
\fbox{\parbox{0.92\textwidth}{\small\setstretch{1.1}
\textbf{Preprint.} This is the authors' submitted version of the article published in
\emph{Journal of Seismology}, 24 (2020), 573--593.
The final version is available at \url{https://doi.org/10.1007/s10950-020-09922-7}.
\textcopyright{} The Author(s) 2020, distributed under a Creative Commons Attribution 4.0 International License (CC BY 4.0).
This version may differ from the published version.}}
\end{center}

\vspace{1em}
\begin{center}
{\Large\bfseries Seismic study of soda straws exposed to nearby blasting vibrations\par}
\vspace{1.2em}
{\large P.~Bottelin$^{1,*}$, L.~Baillet$^{2}$, A.~Mathy$^{3}$, L.~Garnier$^{4}$, H.~Cadet$^{1}$, O.~Brenguier$^{1}$\par}
\vspace{0.8em}
{\small\setstretch{1.1}
$^{1}$ADRGT, 2 rue de la Condamine, 38610 Gières, France\\
$^{2}$ISTerre, CNRS, Université Grenoble Alpes, 38000 Grenoble, France\\
$^{3}$Société Alpine de Géotechnique, 2 rue de la Condamine, 38610 Gières, France\\
$^{4}$SARL Site de la Grotte de Choranche, 2865 route des Grottes, BP 7, 38680 Choranche, France\\
$^{*}$Corresponding author: p.bottelin@adrgt.org\par}
\end{center}

\section*{Abstract}
This research study has been developed to preserve a remarkable soda straw population from vibrations emitted by nearby rock blasting. The Choranche stalactite cave (Vercors, France) contains thousands of exceptionally long soda straws reaching a few meters in length for $\sim$0.5~cm in diameter. These slender speleothems are very vulnerable to vibrations not only because of their structural fragility but also because of their dynamic amplification. We found that soda straws' first natural frequencies lie within the frequency range emitted by nearby rock blasting works (ten to hundreds of Hz). We used in situ blast records, laboratory characterization, and a dynamic 2D finite element code to simulate the load experienced by the soda straw population. We show that induced loads may be increased by a factor 5 due to resonance. Consequently, short soda straws (0.1--1~m) were found more vulnerable to nearby blasting vibrations than longer speleothems ($>$1~m), despite greater own weight and inertial forces for the latter. Simulations made on several blast tests yielded an admissible 2.4~mm/s peak particle velocity along the cave. Subsequent blasting works were carried out in compliance with this threshold with no harm done to the soda straw population. We also computed the dynamic response of soda straws exposed to a regional earthquake. In this case, longer speleothems ($>$1~m) are primarily affected by resonance because of ground motion lower frequency content. We show that the omission of dynamic resonance or its simplification as made in previous studies may significantly underestimate the induced load in speleothems.

\medskip
\noindent\textbf{Keywords:} Blasting works; Rock slope cutting; Speleothems; Numerical modeling; Dynamic amplification; Vibration thresholds

\section*{Highlights}
\begin{itemize}
\item Blasting works may be a significant threat to fragile environmental structures such as soda straws
\item Speleothems represent a natural heritage that must be protected during civil engineering works
\item Risk mitigation studies must take into account the significant dynamic amplification experienced by speleothems
\item Equivalent static computation or single degree of freedom approaches used in previous studies may significantly underestimate the loads in speleothems
\end{itemize}

\section{Introduction}

Due to urban growth, renewal, and densification, geotechnical working sites tend to come closer and closer to natural or human preexisting structures. The use of explosives and heavy machinery for civil engineering works such as quarrying, tunneling, or soil compaction generates potentially harmful vibrations at short to medium distances (tens to hundreds of meters; Crispim 1999; Lacave et al. 2003). This is becoming a major issue for projects taking place in dense areas and underlines the need to reduce vibrations as much as possible. Blast-induced ground vibration level may be contained by several techniques. Trenches and holes, precutting blasts may be used to reduce vibration propagation. The explosive charge per blast hole may be reduced by dividing the bench, and delay time may be used between holes to avoid seismic wave superimposition (Mogi et al. 2000). This requires controlling the firing time of the explosive charge placed in each drilled hole, with the use of precise electrical detonators fired successively with sufficient delay (Hoshino et al. 2000; Rustan 1998). For tunneling or quarrying, numerical softwares are more and more frequently used for blasting excavation optimization (Xie et al. 2018).

In many sedimentary regions, concretionary caves represent a major natural heritage to be preserved (Lacave et al. 2003). It yet also appears to be a vibration-sensitive target (Gauchon et al. 2007). The irreversibility of damages caused to speleothems and the uniqueness of each cave site underlines the need for dedicated vibration studies. Until now, the effects of vibrations on speleothems have been mostly addressed in speleoseismological surveys. Speleoseismology consists in detecting and characterizing speleothem anomalies in caves to study earthquake records (Becker et al. 2006). Growth anomalies, deformation, and/or speleothem failure are notably used to derive some earthquake parameters (e.g. magnitude, azimuth, location, date, peak ground acceleration---PGA or peak ground velocity---PGV) (Bokelmann and Gribovszki 2015; Forti 2001; Forti and Postpischl 1980; Gilli et al. 1999; Méjean et al. 2015; Postpischl et al. 1991; Ferranti et al. 2019; Kagan et al. 2017). The accuracy and robustness of the method yet remains debated as most speleothems---except long and slender ones---may withstand earthquakes (Gilli 2005; Lacave et al. 2004). In addition, many other processes can break speleothems, such as erosion and soil creep, underground glaciers and ice creep, floods, mud and debris flows, frost action, decompression, load and slope movements, and anthropogenic causes (Becker et al. 2006).

Precise characterization of speleothem mechanical properties is of primary importance in speleoseismological studies (Konečný et al. 2015). In the literature, static parameters (geometry, density $\rho$, ultimate strength) as well as elastic and dynamic properties (Young modulus $E$, natural frequencies $f_n$, damping $\xi$) have been provided by both in situ measurements and laboratory experiments (Table~\ref{tab:lit}). Few studies only measured the speleothem density, generally through broken samples weighing and volume estimation. Ultimate compressional ($\sigma_{u,c}$) and tensile ($\sut$) strengths were estimated via uniaxial compression or traction tests, Brazilian tests, or bending tests. Dynamic properties have been measured in situ via geophones or laser interferometers. Significant wandering in mechanical parameters has been observed, both between speleothems of the same site as well as between sites (Bednárik 2009). Such variability may originate from differences in crystallization, porosity, boundary conditions, and/or previous mechanical damage (Bednárik 2009; Szeidovitz et al. 2008). A probabilistic approach has been used by Lacave et al. (2004) to describe variations in tensile rupture stress ($\sut$) between samples.

Due to their slenderness, speleothems may be subject to significant dynamic amplification phenomena. The dynamic behavior of speleothems has generally been addressed in the literature using an equivalent static approach (ESA) (Cadorin et al. 2001; Gribovszki et al. 2008, 2013a, 2013b; Konečný et al. 2015, see Table~\ref{tab:lit}). The speleothem is considered as a Euler-Bernoulli beam loaded at mid-span with a force $F$. This force corresponds to the inertial solicitation, taken equal to the speleothem mass multiplied by the peak ground acceleration (PGA). In ESA, the maximal tensile stress is located at the embedded extremity. The ground horizontal acceleration causing the rupture is then only function of the speleothem geometry (length $L$, internal and external diameters $D_{int}$ and $D_{ext}$), density $\rho$, and ultimate strength $\sigma_u$. For such loading, the speleothem bending rupture depends on the calcite tensile strength $\sut$ rather than its compressional resistance $\sigma_{u,c}$.

A generalization to irregular beams was proposed by Lacave et al. (2004) and Cadorin et al. (2001). However, the dynamic resonance that occurs when the ground motion frequency content matches the speleothem natural frequencies cannot be simulated in the ESA. This technique is hence restricted to the ``rigid-body'' domain, i.e. when the ground motion occurs at lower frequencies than speleothem natural frequencies (the latter generally ranging between tens of Hz to hundreds of Hz). Such conditions are not systematically fulfilled: slender speleothems may indeed experience significant amplification phenomena under earthquakes, allowing rupture at lower acceleration values (Gribovszki et al. 2017; Lacave et al. 2004). Long, slender, tubular speleothems such as soda straws are thus particularly vulnerable to ground motion, making them suitable targets for speleoseismological studies (Becker et al. 2006; Lacave et al. 2000).

In their speleoseismological study, Lacave et al. (2004) adapted Eurocode 8 elastic response spectra. For speleothems with fundamental natural frequency $f_0$ lying in the seismic bandwidth ($<$25~Hz), a 4.5 flat amplification factor was applied. This plateau amplification may yet be conservative or unconservative, depending on speleothem mechanical properties as well as ground motion spectrum. In addition, previous studies used only the peak ground acceleration (PGA) to compute the maximal bending stress undergone by speleothems (Cadorin et al. 2001; Gribovszki et al. 2017; Lacave et al. 2004). In contrast, the acceleration full waveform has never been used, although it contains the whole ground motion information.

To our knowledge, the study of vibrations induced by geotechnical engineering works located close to a concretionary cave has been addressed only by Lacave et al. (2003). Their approach yet does not simulate entirely the dynamic amplification that speleothems may experience. Our study focuses on blasting excavation works conducted to restore the pedestrian access to the Choranche cave (Vercors, France). The rock blasting takes place only tens of meters away to very delicate, hollow tubular soda straw speleothems. We first characterize soda straw dynamic parameters and ultimate strength thanks to laboratory and in situ experiments. We then carry out blast tests to record ground motion along the cave wall and characterize the wave attenuation with distance. We use a dynamic finite element modeling (FEM) to estimate the state of stress in the soda straws subjected to the recorded ground motion. Finally, the admissible peak particle velocity (PPV) along the cave wall was estimated, as well as the maximal instantaneous charge of explosive (MIC) as a function of distance.

\begin{landscape}
\footnotesize
\setlength{\tabcolsep}{4pt}
\renewcommand{\arraystretch}{1.15}
\begin{longtable}{L{2.1cm}L{3.5cm}L{5.2cm}L{1.9cm}L{2.0cm}L{1.5cm}L{1.9cm}L{1.6cm}}
\caption{Literature review of speleothem mechanical properties (part 1: references, studied speleothems and laboratory experiments).}\label{tab:lit}\\
\toprule
Reference & Cave & Speleothem type & \multicolumn{2}{l}{Density $\rho$ (kg\,m$^{-3}$)} & \multicolumn{2}{l}{Ultimate tensile strength $\sut$ (MPa)} & Young's modulus $E$ (GPa) \\
\cmidrule(lr){4-5}\cmidrule(lr){6-7}\cmidrule(l){8-8}
 & & & Experiment & Value & Experiment & Value & Experiment \\
\midrule
\endfirsthead
\multicolumn{8}{l}{\textbf{Table \thetable} (continued)}\\
\toprule
Reference & Cave & Speleothem type & \multicolumn{2}{l}{Density $\rho$ (kg\,m$^{-3}$)} & \multicolumn{2}{l}{Ultimate tensile strength $\sut$ (MPa)} & Young's modulus $E$ (GPa) \\
\cmidrule(lr){4-5}\cmidrule(lr){6-7}\cmidrule(l){8-8}
 & & & Experiment & Value & Experiment & Value & Experiment \\
\midrule
\endhead
\bottomrule
\endfoot
Gribovszki et al. 2008 & Snezanka and Eminova, Bulgaria & Stalagmites ($1.1 < L < 1.40$~m, $3.0 < d < 13.7$~cm) & Not described & $2230 \pm 0.15$ & BrT & $2.61 \pm 0.28$ & UCS \\
Gribovszki et al. 2013a & Baradla, Hungary \newline Hajnóczy, Hungary \newline Domica, Slovakia \newline Varteshkata, Bulgaria \newline Elata, Bulgaria \newline Snezanka, Bulgaria \newline Eminova, Bulgaria & Stalagmites ($1.17 < L < 5.10$~m, $3.0 < d < 13.7$~cm) & Not described & $2390 \pm 160$ \newline -- \newline $2350 \pm 120$ \newline 2400 \newline $2500 \pm 80$ \newline $2320 \pm 150$ \newline $2140 \pm 250$ & BrT & $1.62 \pm 0.48$ \newline -- \newline $2.75 \pm 0.56$ \newline $1.62 \pm 0.42$ \newline $1.62 \pm 0.42$ \newline $2.61 \pm 0.28$ \newline $3.83 \pm 0.81$ & UM \\
Gribovszki et al. 2013b & Domica, Slovakia & Stalagmites ($1.8 < L < 5.0$~m, $4.0 < d < 11.0$~cm) & Not described & $2357.85 \pm 115.8$ & BrT & $2.64 \pm 0.56$ & UCS; UM \\
Gribovszki et al. 2017 & Plavecká & Stalagmite ($L = 4.3$~m, $7.0 < d < 10.5$~cm) & Not described & $1940.5 \pm 6.4$ & DT & $0.51 \pm 0.13$ & UM \\
Cadorin et al. 2001 & Hotton, Belgium & Speleothems ($0.06 < L < 1.0$~m, $3.0 < d < 27.0$~cm) & None & 2500 & BrT, BT (3PFT) & $1.40 \pm 0.77$ & None \\
Konečný et al. 2015 & Baradla-Domica, Hungary \& Slovakia & Speleothem fragments ($d \sim 5$~cm) & Water displacement \& X-ray tomography & 2289 & -- & -- & UM \\
Lacave et al. 2000 & Choranche, France & 6 stalagmites ($0.34 < L < 0.82$~m, $4.0 < d < 12.0$~cm), 3 soda straws ($0.09 < L < 0.28$~m, $0.5 < d < 0.8$~cm), 3 stalactites ($0.25 < L < 0.45$~m, $0.8 < d < 5.0$~cm) & -- & 2600 (from Kourimsky and Tvrz, 1983) & -- & -- & $f_n$ \\
 & Antre de Vénus, France & 2 soda straws ($0.7 < L < 0.8$~m, $0.5 < d < 1.7$~cm), 4 stalactites ($0.64 < L < 0.89$~m, $3.2 < d < 4.0$~cm), 1 giant stalactite ($L \sim 3.5$~m, $30 < d < 60.0$~cm) and 1 column ($L \sim 0.78$~m, $d \sim 6.0$~cm) & -- & & -- & -- & \\
Lacave et al. 2003, 2004 & Milandre, Switzerland & Soda straws, stalactites, stalagmites ($0.1 < L < 10.0$~m, $0.5 < d < 10.0$~cm) & -- & 2600 (from Kourimsky and Tvrz, 1983) & BT (EF) & $4.26 \pm 1.73$ & -- \\
Szeidovitz et al. 2008 & Hajnóczy, Hungary \newline Baradla, Hungary & Stalagmite ($L \sim 3.6$~m, $d \sim 5.0$~cm) \newline Stalagmites and pillar ($2.1 < L < 5.1$~m, $4.5 < d < 29.0$~cm) & -- \newline Not described & -- \newline $2394 \pm 155$ & -- \newline BrT & -- \newline $1.62 \pm 0.48$ & -- \newline UCS \\
Gilli et al. 1999 & Saint-Paul-de-Fenouillet, France & Soda straw fragments ($d \sim 0.5$~cm) & -- & -- & Not described & 0.93--2.8 & Rough estimate (no test) \\
This study & Choranche, France & Soda straws ($0.025 < L < 3.0$~m, $d \sim 0.45$~cm) & Water displacement & 2500 & BT (EF \& 3PFT) & 2.5--16.2 (most of values in the range 2.5--7.5) & $f_n$ \\
\end{longtable}

\begin{longtable}{L{3.6cm}L{2.6cm}L{5.2cm}L{4.0cm}L{2.6cm}L{1.4cm}L{2.2cm}}
\caption*{\textbf{Table \ref{tab:lit}} (continued; part 2: laboratory Young's modulus and damping, in situ experiments and modeling).}\\
\toprule
Reference & \multicolumn{2}{l}{Laboratory experiments} & \multicolumn{2}{l}{In situ experiments} & \multicolumn{2}{l}{Modeling} \\
\cmidrule(lr){2-3}\cmidrule(lr){4-5}\cmidrule(l){6-7}
 & Young's modulus $E$ (GPa) & Damping $\xi$ (\%) & Sensors & Speleothem parameters & Method & Amplification \\
\midrule
\endfirsthead
\toprule
Reference & \multicolumn{2}{l}{Laboratory experiments} & \multicolumn{2}{l}{In situ experiments} & \multicolumn{2}{l}{Modeling} \\
\cmidrule(lr){2-3}\cmidrule(lr){4-5}\cmidrule(l){6-7}
 & Young's modulus $E$ (GPa) & Damping $\xi$ (\%) & Sensors & Speleothem parameters & Method & Amplification \\
\midrule
\endhead
\bottomrule
\multicolumn{7}{p{22cm}}{Reference, cave name, and type of speleothem are given in the first columns. When available, length $L$ and diameter $d$ of the studied speleothems are given. For laboratory experiments, measurement method and parameter value (mean and standard deviation) are given, when available. The ultimate tensile strength $\sut$ has been measured in the lab via Brazilian tests (BrT), direct traction test (DT), or bending tests (BT). Bending tests can be embedded-free (EF) or three-point flexural tests (3PFT). Young's modulus $E$ has been characterized by uniaxial compressive strength tests (UCS), ultrasonic velocity (UM), or measurement of natural frequencies ($f_n$). Damping $\xi$ is expressed as percentage relative to critical damping. For in situ experiments, we indicate the speleothem parameters, which were measured ($f_n$: natural frequencies, $\xi$: damping, $E$: Young's modulus). Equivalent static approach (ESA) based on Euler-Bernoulli beam theoretical formulas, response spectra (RS), and finite element modeling (FEM) methods have been used to describe the speleothem's response to ground motion. N/A: not taken into account.}\\
\endlastfoot
Gribovszki et al. 2008 & 6.24 & -- & Geophones & $f_n$ & ESA & N/A \\
Gribovszki et al. 2013a & $20.8 \pm 5.9$ \newline -- \newline $23.6 \pm 4.0$ \newline 53 \newline -- \newline -- \newline -- & -- & Geophones & $f_n$ & ESA & N/A \\
Gribovszki et al. 2013b & $23.4 \pm 4.4$ & -- & Geophones & $f_n$ & ESA & N/A \\
Gribovszki et al. 2017 & $25.2 \pm 3.92$ & -- & Geophones & $f_n$ & ESA & N/A \\
Cadorin et al. 2001 & 22 & -- & -- & -- & ESA & N/A \\
Konečný et al. 2015 & 33.8 & -- & -- & -- & -- & -- \\
Lacave et al. 2000 & 37--72 \newline 28--47 & 2.3 (all speleothems); 0.2 (soda straws only) \newline 0.12 (all speleothems); 0.08 (soda straw only) & Laser interferometer & $f_n$, $\xi$, $E$ & ESA & N/A \\
Lacave et al. 2003, 2004 & -- & 0.1 (from Lacave et al. 2000) & Accelerometers & -- & RS & Arbitrary factor \\
Szeidovitz et al. 2008 & -- \newline $20.8 \pm 5.9$ & -- & Geophones & $f_n$, $\xi$ & ESA & N/A \\
Gilli et al. 1999 & 0.4 & -- & -- & -- & -- & -- \\
This study & 60 & 0.2--2.0 & Geophones, Laser interferometer & $f_n$, $\xi$, $E$ & FEM & Yes \\
\end{longtable}
\addtocounter{table}{-1}
\end{landscape}

\section{Site description}
\label{sec:site}

\begin{figure}[htbp]
\centering
\includegraphics[width=0.85\textwidth]{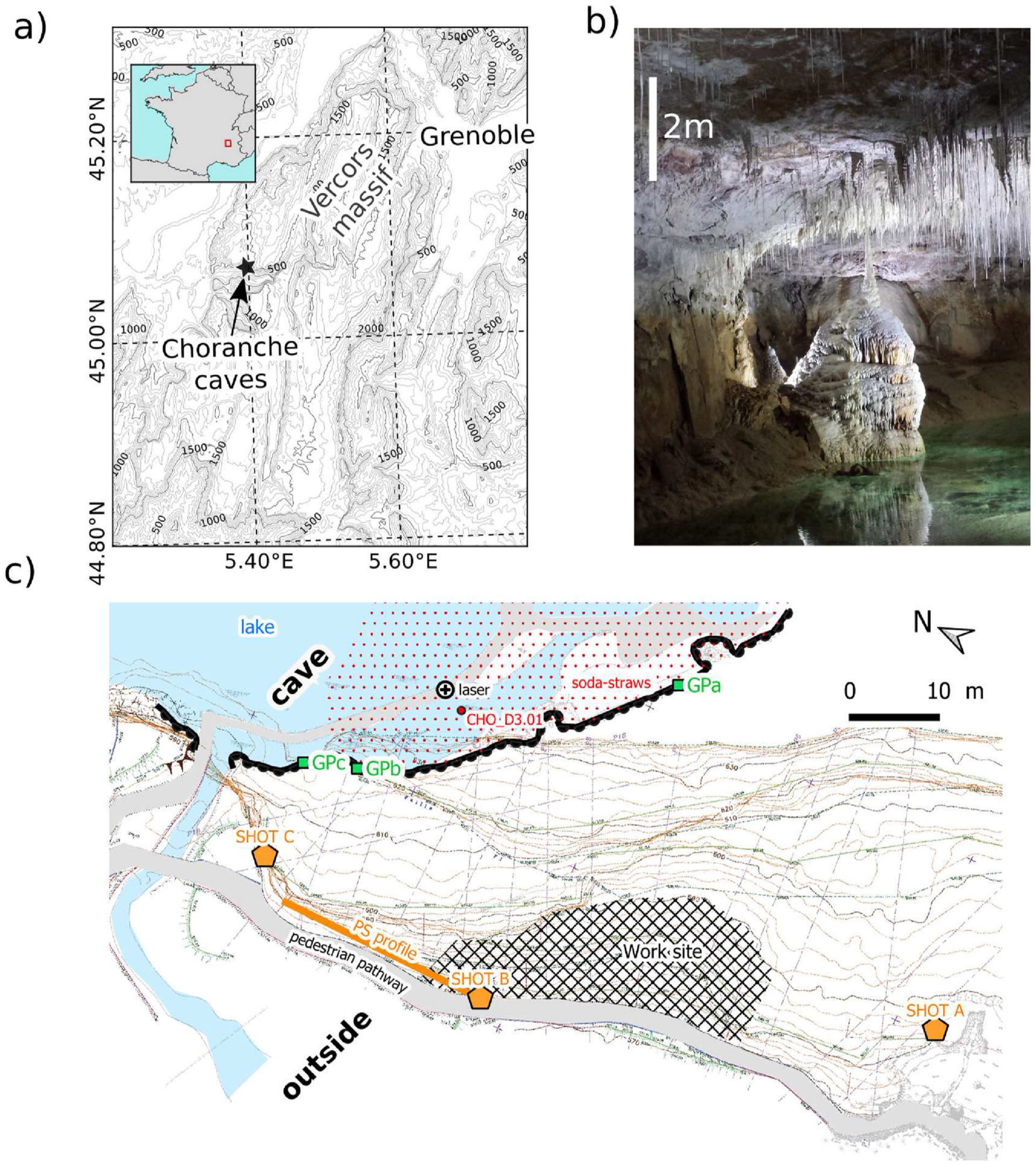}
\caption{a) Location map. Choranche site is shown with a black star. b) Picture taken inside the cave (lake room). Soda straw speleothems range mostly between 1 to 3~m in the picture. c) Map of the site. The cave wall is delineated with the thick black line, the rounded side pointing toward the rock mass. The rock mass is shown in white. The lake is shown in blue. Pedestrian pathways are shown in light gray. The soda straws (red dotted area) are located about tens of meters away from the blasting excavation work site (black hatched area). Seismic sensors (GPa, GPb, GPc) installed on the internal cave wall are shown with green squares. Tests shot locations (A, B, C) are shown with orange pentagons. The seismic profile (PS) is shown with the orange segment.}
\label{fig:site}
\end{figure}

The study site is located in the Isère department (southeast France), in the Vercors Pre-Alps Mesozoic massif (Fig.~\ref{fig:site}a). The Choranche cave consists of two main rooms (lake and cathedral, Fig.~\ref{fig:site}b and c) and several halls and galleries. The cave is included in a $>$50~km long karstic network called Gournier-Coufin-Chevaline (Delannoy 1986) and located about 350~m below the surface. This network develops at the interface between massive Urgonian limestone (lower Cretaceous) and underlying impervious marls and calcareous marls from Hauterivian (Delannoy et al. 2009). The cave hosts many speleothems, including very long (up to 3~m), thin, translucent soda straws (Fig.~\ref{fig:site}b). These hollow cylinders get longer via successive crystallization of calcite rings around water drops hanging at the speleothem's end (Fairchild and Baker 2012). The mean annual growth rate ranges between 0.09 and 0.92~mm/year (Perrette and Jaillet 2010) so that the longest Choranche soda straws may date from 3200 to 33,000 BP. Most of the soda straws concentrate in the lake room and the soda straw hall close to the cave exit (Fig.~\ref{fig:site}b and c).

This concretional cave represents an exceptional geological inheritance (Delannoy et al. 2009; Gauchon et al. 2007; Reynard et al. 2011). The cave site is opened to the public throughout the year and welcomes about 100,000 visitors per year (Garnier, pers. comm. 2018). A narrow, hanging pedestrian pathway located at the cliff toe gives access to the cave (Fig.~\ref{fig:site}c). Rehabilitation works were required, consisting in moving the pedestrian pathway closer to the rock wall along about 100~m. As the rock site is not accessible for heavy machinery, explosives were planned for rock slope cutting.

Two drill cores (15 and 50~m) and a ground penetrating radar (GPR) campaign were carried out during fall 2017 and winter 2017--2018 (not shown here). They revealed a major subvertical fracture (N175/subvertical) striking roughly parallel to both the outward cliff and the cave wall (Fig.~\ref{fig:site}c). The rock mass at depth appeared massive, with a rock quality designation (RQD) $>$90\% and no significant reflector. In contrast, the external rock skin (3 to 6~m in depth) showed intense fracturing. Such rock mass decoupling and fracturing was expected to significantly reduce the ground motion level into the cave (Kumar et al. 2016). The distance between the blasts and the soda straw ranges about tens of meters (Fig.~\ref{fig:site}c).

\section{Soda straw characterization}

\subsection{Dynamic properties}
\label{sec:dyn}

Fifteen samples of soda straws were collected into the cave, mostly found lying onto the lake bottom (dataset D1). We measured their length, diameter, and wall thickness using a Vernier caliper. Most of the soda straws show a perfectly regular, cylindrical shape, although a few of them contain enlarged cross-sections. The length of samples ranges between a few centimeters up to about 40~cm. Longer soda straws cannot be found intact in the cave as they are probably smashed into small bits when impacting the ground or the lake. The mean external diameter $D_{ext}$ for regular-shaped soda straws is $4.5 \times 10^{-3}$~m and average calcite wall thickness is $5 \times 10^{-4}$~m.

The speleothem density was characterized for a larger speleothem sample coming from the same Choranche cave. The speleothem density $\rho = 2500$~kg\,m$^{-3}$ was computed from sample weighing with a scale and volume estimation from water displacement method. This value falls in the upper range compared with previous studies, probably because of lower porosity in soda straws. It agrees very well with values taken for soda straws by Lacave et al. (2000).

\begin{figure}[htbp]
\centering
\includegraphics[width=\textwidth]{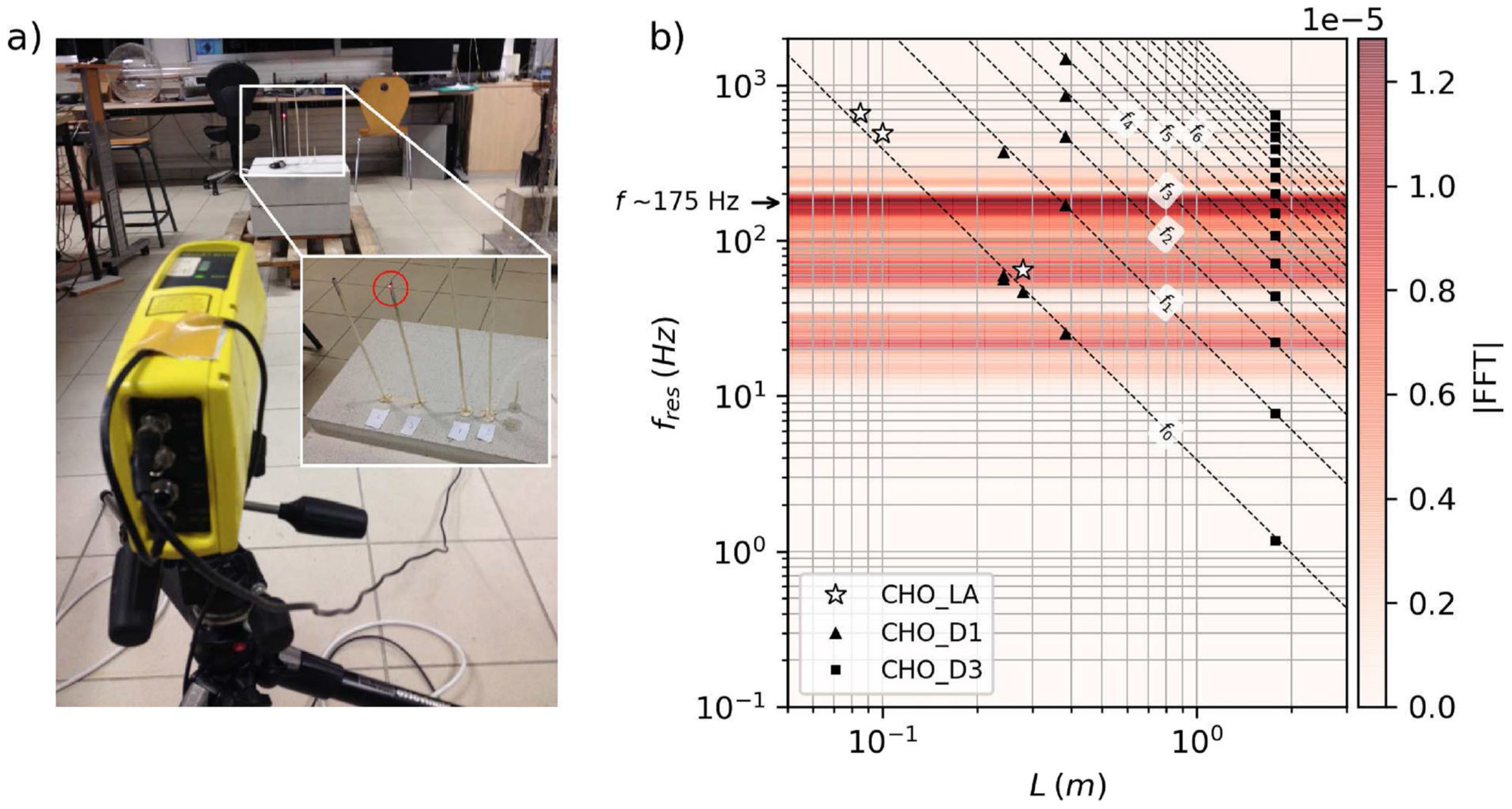}
\caption{a) Laboratory experimental setup for soda straw dynamic characterization. The laser vibrometer is visible at the forefront, pointing at the speleothem's top (see red laser dot in the zoom inset, delineated with the red circle). Soda straws CHO\_D1.01, CHO\_D1.02, CHO\_D1.06, and CHO\_D1.07 are sealed vertically with chemical grout into a concrete block (background). b) Soda straw resonant frequencies ($f_0$--$f_{13}$) as a function of their length $L$. Soda straws from datasets CHO\_D1 (D1.01, D1.02, D1.06, and D1.07) and CHO\_D3 are shown with black triangles and black squares, respectively (this study). Measurements from Lacave et al. (2000) are shown with white stars (CHO\_LA). Theoretical resonant frequencies ($f_0$--$f_{13}$) from Eq.~\ref{eq:fn} are shown as dashed black lines and labeled for $f_0$--$f_6$. The background red color scale shows the spectral content of shot A (see Sect.~\ref{sec:seismic}), with a clear spectral peak close to 175~Hz.}
\label{fig:lab}
\end{figure}

We carried out dynamic testing on 4 soda straw samples in the laboratory (dataset CHO\_D1) and on 1 soda straw in situ (dataset CHO\_D3, Table~\ref{tab:dyn}). The length of CHO\_D3.01 could not be measured on the field because of many surrounding, fragile soda straw preventing access to the cave ceiling. Due to instrument availability issues, we used an Ometron VH300+ Laser Doppler Vibrometer Type 8329 (\url{www.bksv.com}) for the lab experiment and a Polytec PDV-100 (\url{www.polytec.com}) for the in situ measurements. Figure~\ref{fig:lab}a shows the laboratory experimental setup. The laser beam points at the speleothem top, which was painted in metallic silver for better reflection coefficient. For the in situ experiment, the water dripping on the soda straw caused proper laser reflection and a good signal-to-noise ratio. The bottom of the speleothem was embedded into a concrete block with a chemical seal. The speleothem was excited either with compressed air or light hammer impacts on the concrete block. The velocity of the speleothem extremity was recorded during 10~s at 5000~Hz. For the in situ instrumentation, the laser was setup on a pole to reach the height of the soda straw free extremity. The pole position was strongly constrained by the location of the pedestrian pathway in the cave, the lake preventing the instrument setup in the area (Fig.~\ref{fig:site}c). In addition, it had to be installed without causing any harm to the soda straws, which restricted again the eligible areas. The pole natural frequencies were measured by pointing the laser beam against a wall and were removed from the records. For each shot, the signal was sampled at 2000~Hz.

For both laboratory and field experiment, the natural frequencies ($f_n$) were pointed on the laser vibrometer record spectra (Table~\ref{tab:dyn}). Previous works conducted by Lacave et al. (2000) in the same cave have also been included (dataset CHO\_LA, Table~\ref{tab:dyn}).

The theoretical natural frequencies of a Euler-Bernoulli beam are given by the analytical formula in Eq.~\ref{eq:fn} (from Chopra 2001) and drawn with black dashed lines in Fig.~\ref{fig:lab}b.
\begin{equation}
f_n = \frac{X_n^2}{2\pi}\sqrt{\frac{EI}{\mu L^4}}
\label{eq:fn}
\end{equation}
with:\\[2pt]
\begin{tabular}{@{}ll@{}}
$f_n$ & natural frequency at $n$th mode (Hz); \\
$X_n^2$ & mode coefficient (without unit, see Table~\ref{tab:modes}); \\
$E$ & Young's modulus (N\,m$^{-2}$); \\
$I$ & moment of inertia of the beam (m$^4$), equal to $\pi (D_{ext}^4 - D_{int}^4)/64$ for a hollow cylinder; \\
$\mu$ & mass per unit length (kg\,m$^{-1}$), equal to $\rho (\pi/4)(D_{ext}^2 - D_{int}^2)$ for a hollow cylinder; \\
$L$ & beam length (m). \\
\end{tabular}

\medskip
Figure~\ref{fig:lab}b presents the natural frequencies $f_n$ as a function of speleothem length $L$. Experimental tests for both laboratory experiment (CHO\_D1, black triangles) and in situ measurements (CHO\_LA, white stars) show consistent results for soda straw ranging from a few centimeters up to about 30~cm in length (Fig.~\ref{fig:lab}b). The best fit between theoretical natural frequencies (dashed black lines) and experimental values was obtained for a 60~GPa Young's modulus $E$. This value agrees very well with previous studies conducted on soda straws (Lacave et al. 2000). It is about 2 to 3 times greater than for other kind of speleothems (see Table~\ref{tab:lit}), probably because of complex shape, porosity effects, and/or crystallization defects for the latter.

Natural frequencies of the CHO\_D3.01 soda straw have been measured up to the 13th mode (black squares, Fig.~\ref{fig:lab}b). Experimental resonance frequencies and the ratios between successive modes are both perfectly described by Euler-Bernoulli beam theoretical formulas (Eq.~\ref{eq:fn} and Table~\ref{tab:modes}). We assumed that embedded-free conditions with chemical seal in the laboratory experiment are comparable with calcite bond in caves. We hence estimated the CHO\_D3.01 soda straw length at 1.78~m, which is in good agreement with visual estimate in the field. Our results suggest that Choranche soda straws can be accurately modeled as regular, tubular beams for dynamic characterization.

The damping of soda straw was measured via logarithmic decrement technique (Inman 2008, Table~\ref{tab:dyn}). The damping $\xi$ is expressed in percent relative to critical damping. The damping $\xi_0$ measured at the fundamental frequency is generally very low ($<$2\%), with a mean of about 0.5\%. This value is compatible with measurements from Lacave et al. (2000) carried out on the same Choranche soda straws (same table). Higher modes show damping ranging between about 0.1 and 1\%. In the rest of this study, we used a $\xi_d = 0.1$\% design damping value for all modes in our computations.

\begin{landscape}
\begin{table}[p]
\centering
\caption{Geometrical characterization and dynamic parameters of Choranche soda straw speleothems.}
\label{tab:dyn}
\footnotesize
\setlength{\tabcolsep}{3.5pt}
\begin{tabular}{llllcccccccccccccc}
\toprule
Sample id & Experiment & $L$ (m) & Parameter & \multicolumn{14}{c}{Mode number} \\
\cmidrule(l){5-18}
 & & & & 0 & 1 & 2 & 3 & 4 & 5 & 6 & 7 & 8 & 9 & 10 & 11 & 12 & 13 \\
\midrule
\emph{CHO\_D1.01} & Lab & 0.38 & $f_n$ (Hz) & 25.6 & 168.0 & 460.0 & 869.0 & 1481.0 & & & & & & & & & \\
 & & & $\xi_n$ (\%) & 0.20 & 0.23 & 1.00 & 2.00 & & & & & & & & & & \\
\emph{CHO\_D1.02} & Lab & 0.28 & $f_n$ (Hz) & 47.3 & & & & & & & & & & & & & \\
 & & & $\xi_n$ (\%) & & & & & & & & & & & & & & \\
\emph{CHO\_D1.06} & Lab & 0.24 & $f_n$ (Hz) & 60.4 & & & & & & & & & & & & & \\
 & & & $\xi_n$ (\%) & & & & & & & & & & & & & & \\
\emph{CHO\_D1.07} & Lab & 0.24 & $f_n$ (Hz) & 57.3 & & & & & & & & & & & & & \\
 & & & $\xi_n$ (\%) & & & & & & & & & & & & & & \\
\emph{CHO\_D3.01} & Field & 1.78$^{\S}$ & $f_n$ (Hz) & 1.2 & 7.8 & 22.4 & 44.0 & 72.1 & 108.4 & 151.0 & 200.9 & 256.2 & 320.3 & 391.0 & 467.0 & 545.0 & 650.0 \\
 & & & $\xi_n$ (\%) & 1.7 & 0.63 & 0.24 & 0.12 & 0.09 & 0.09 & 0.09 & 0.1 & 0.09 & -- & -- & -- & -- & -- \\
 & Numerical & 1.78$^{\S}$ & $f_n$ (Hz) & 1.2 & 7.7 & 21.6 & 42.4 & 70.0 & 104.7 & 146.2 & 194.7 & 250.1 & 312.4 & 381.6 & 457.8 & 540.9 & 630.9 \\
 & & & $\xi_n$ (\%) & \multicolumn{2}{l}{fixed -- 0.1} & & & & & & & & & & & & \\
\rowcolor{gray!15}\emph{CHO\_LA.FIS1} & Field & 0.085 & $f_n$ (Hz) & 666 & & & & & & & & & & & & & \\
\rowcolor{gray!15} & & & $\xi_n$ (\%) & 0.04 & & & & & & & & & & & & & \\
\rowcolor{gray!15}\emph{CHO\_LA.FIS2} & Field & 0.1 & $f_n$ (Hz) & 492 & & & & & & & & & & & & & \\
\rowcolor{gray!15} & & & $\xi_n$ (\%) & 0.11 & & & & & & & & & & & & & \\
\rowcolor{gray!15}\emph{CHO\_LA.FIS3} & Field & 0.28 & $f_n$ (Hz) & 65 & 448 & & & & & & & & & & & & \\
\rowcolor{gray!15} & & & $\xi_n$ (\%) & 0.42 & & & & & & & & & & & & & \\
\bottomrule
\end{tabular}

\smallskip
\parbox{0.95\linewidth}{\footnotesize \emph{CHO\_D1} and \emph{D3}: datasets measured in this study; \emph{CHO\_LA}: dataset from Lacave et al. (2000) study (gray backfill). Boundary conditions correspond to embedded-free tests. Speleothem length $L$ is indicated in the table. Natural frequency $f_n$ at the $n$th natural mode is given along with corresponding damping ratio $\xi_n$ (in percent relative to critical damping). $^{\S}$Estimated from natural frequencies.}
\end{table}
\end{landscape}

\begin{table}[htbp]
\centering
\caption{Mode coefficients $X_n^2$ for Euler-Bernoulli beam natural frequencies.}
\label{tab:modes}
\begin{tabular}{lcccc}
\toprule
Mode number & 0 & 1 & 2 & 3 and higher \\
\midrule
Coefficient $X_n^2$ & $1.875102^2$ & $4.694092^2$ & $7.854732^2$ & $\left((2n+1)\,\pi/2\right)^2$ \\
\bottomrule
\end{tabular}
\end{table}

\subsection{Rupture parameters}
\label{sec:rupture}

Soda straw failure parameter (i.e. ultimate tensile strength $\sut$) was measured on another set of samples (dataset D2, Table~\ref{tab:fail}). We used both single-end loading on embedded-free speleothems (EF) and three-point flexural test (3PFT, Table~\ref{tab:lit}). The load $P$ was progressively increased by pouring sand in a recipient attached to the speleothem extremity (EF) or mid-span (3PFT). The tensile stress at failure was computed using the formula given in Eq.~\ref{eq:sigma} with $K = 32$ (EF) and $K = 8$ (3PFT), respectively.
\begin{equation}
\sut = \frac{K\,P\,L\,D_{ext}}{\pi\left(D_{ext}^4 - D_{int}^4\right)}
\label{eq:sigma}
\end{equation}
The speleothem own-weight was not measured during computation of $\sut$ as it is negligible compared with the applied load $P$.

Our results (Table~\ref{tab:fail}) show that $\sut$ wanders greatly between 2.5 and 16.2~MPa, most of values concentrating between 2.5 and 7.5~MPa. The lowest values might be caused by preexisting microfractures because the collected samples have experienced failure: they were found lying on the ground or on the bottom of the lake. Tensile strength $\sut$ measured agree well with previous studies conducted on soda straws (Table~\ref{tab:lit}). Other types of speleothems generally show lower tensile strength because of porosity effects, crystallization defects, and/or microdamage.

In the rest of this article, we selected the characteristic value of failure strength as the weakest value of our tests, i.e. about 2.5~MPa. We then defined a conservative design value $\sutd = 1.25$~MPa so that soda straws would not endure more than $\sim$50\% of their ultimate failure. This value corresponds to a $\sim$5\% rupture probability based on the tests conducted in Milandre cave (Lacave et al. 2003).

\begin{table}[htbp]
\centering
\caption{Results of failure tests conducted on dataset CHO\_D2. B.C.: boundary conditions.}
\label{tab:fail}
\small
\begin{tabular}{llc@{\hspace{2.5em}}llc}
\toprule
Failure test & B.C. & $\sut$ (MPa) & Failure test & B.C. & $\sut$ (MPa) \\
\midrule
CHO\_D2.01 & EF & 5.61 & CHO\_D2.24 & 3PFT & 4.48 \\
CHO\_D2.02 & EF & 4.72 & CHO\_D2.25 & 3PFT & 3.46 \\
CHO\_D2.03 & EF & 3.54 & CHO\_D2.26 & 3PFT & 3.52 \\
CHO\_D2.04 & EF & 12.02 & CHO\_D2.27 & 3PFT & 4.56 \\
CHO\_D2.05 & EF & 8.40 & CHO\_D2.28 & 3PFT & 4.66 \\
CHO\_D2.06 & EF & 13.56 & CHO\_D2.29 & 3PFT & 4.07 \\
CHO\_D2.07 & EF & 10.12 & CHO\_D2.30 & 3PFT & 5.36 \\
CHO\_D2.08 & EF & 9.40 & CHO\_D2.31 & 3PFT & 2.98 \\
CHO\_D2.09 & EF & 16.21 & CHO\_D2.32 & 3PFT & 4.15 \\
CHO\_D2.10 & EF & 9.10 & CHO\_D2.33 & 3PFT & 6.95 \\
CHO\_D2.21 & 3PFT & 5.83 & CHO\_D2.34 & 3PFT & 9.94 \\
CHO\_D2.22 & 3PFT & 3.16 & CHO\_D2.35 & 3PFT & 14.89 \\
CHO\_D2.23 & 3PFT & 2.45 & & & \\
\midrule
\multicolumn{2}{l}{Min (MPa)} & 2.45 & \multicolumn{2}{l}{Median (MPa)} & 5.36 \\
\multicolumn{2}{l}{Max (MPa)} & 16.21 & \multicolumn{2}{l}{Standard deviation (MPa)} & 3.98 \\
\multicolumn{2}{l}{Mean (MPa)} & 6.93 & \multicolumn{2}{l}{Design value $\sutd$ (MPa)} & 1.25 \\
\bottomrule
\end{tabular}

\smallskip
\parbox{0.8\textwidth}{\footnotesize In three-point flexural test (3PFT), the speleothem lies on two supports with loading at mid-span. EF: embedded-free with end-loading.}
\end{table}

\section{Seismic measurements}
\label{sec:seismic}

We measured the ground motion velocity in three locations (GPa, GPb, GPc, Fig.~\ref{fig:site}c) spread out along the cave wall. Three orthogonal 1C, 4.5~Hz geophones were set at each location. Vertical (Z) and transverse (T) sensors were mounted on metallic L-brackets. Geophone coupling with the rock was achieved by pouring plaster into the drill holes. Radial component (R) was facing the outward normal to the cave wall. We used a Geometrics\textsuperscript{\textregistered} Geode\textsuperscript{\textregistered} acquisition system operating at 4000~Hz during 4~s. The records started at the same time as source triggering.

Blast tests were conducted using Titanobel\textsuperscript{\textregistered} Emulstar 8000 UG and Daveydet\textsuperscript{\textregistered} electrical detonators. The maximal instantaneous charge (MIC) ranged from 0.013 to 0.2~kg. The shots were triggered individually at one go. Explosive was fired in holes drilled into the rock at three different spots (A, B, C, Fig.~\ref{fig:site}c). Each firing hole was used only one time.

\begin{figure}[htbp]
\centering
\includegraphics[width=\textwidth]{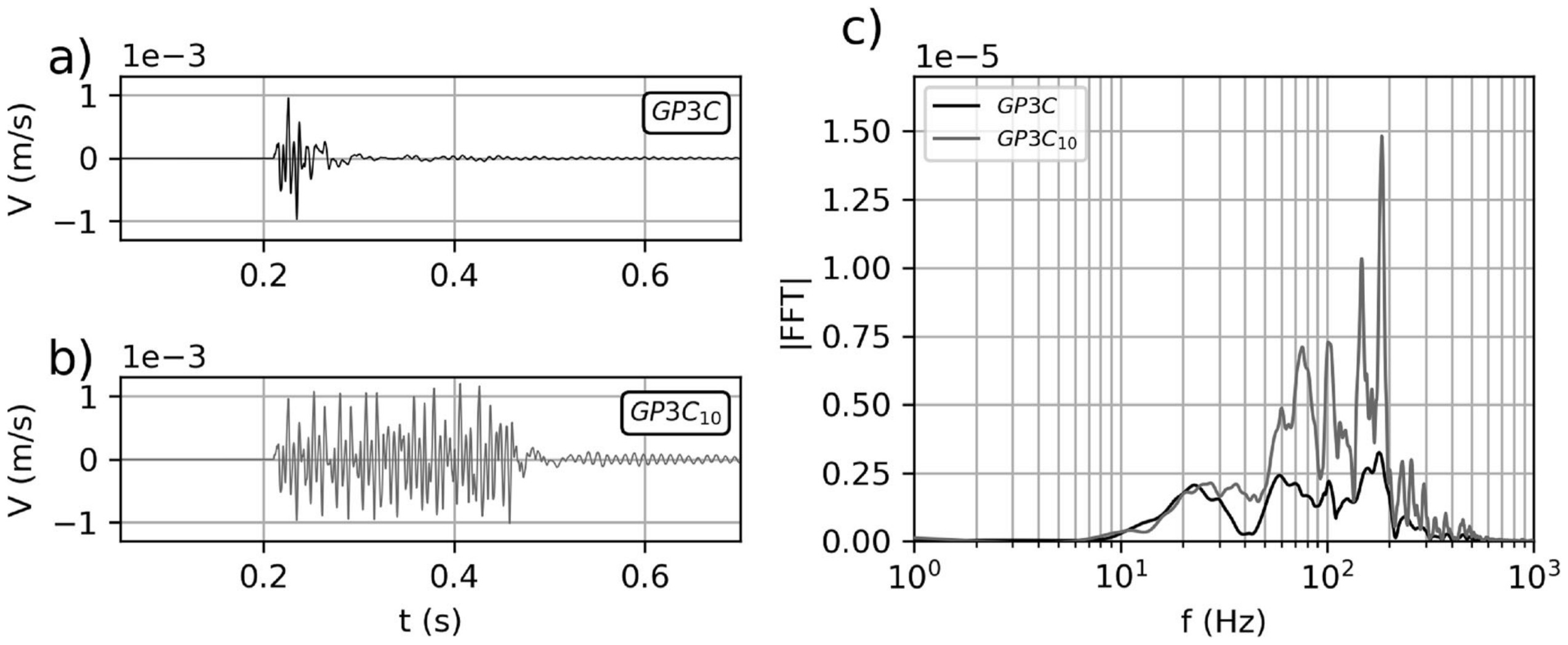}
\caption{a, b) Ground particle velocity for signals \GPC{} and \GPCten, respectively. Signal \GPC{} derives from test blast A (MIC = 0.2~kg) recorded at sensor GPa inside the cave ($D \approx 55$~m). Synthetic signal \GPCten{} corresponds to 10 micro delays randomly distributed between 20 to 30~ms. c) Fourier spectra of \GPC{} (black) and \GPCten{} (gray).}
\label{fig:signal}
\end{figure}

The ground motion recorded at sensor GPa during blast test A (MIC = 0.2~kg, $d \approx 55$~m) is shown in Fig.~\ref{fig:signal}a. Due to instrument malfunction, only the radial component was usable. The signal amplitude was multiplied by $\sqrt{3}$ as a conservative proxy of dominant motion (Chapot 1981a). This hypothesis was later confirmed during excavation blasts. The signal shown in Fig.~\ref{fig:signal}a is referred as \GPC{} in the rest of this study. The velocity signal shows a sudden burst in ground velocity, reaching a peak particle velocity (PPV) of about 1~mm/s. The vibration then decays rapidly after about 0.1~s in length. \GPC{} spectral content (Fig.~\ref{fig:signal}c, black curve) shows a wide, regular spectrum, spreading between 10 and about 500~Hz. The spectrum culminates between 150 and 200~Hz and shows negligible amplitude below 10~Hz and over 500~Hz. The signal shape and spectrum are both consistent with an explosive source fired at short-medium distance (tens of meters) and traveling through a rock medium.

\section{Speleothem vibration modeling}

The Choranche soda straws were modeled as bending beams, without considering the shear deformation and rotational inertia. A homemade elastodynamic 2D code with Euler-Bernoulli finite elements is used to simulate either a modal analysis (computation of mode shapes and natural frequencies) or a transient response analysis (temporal evolution of displacements, velocities, and stresses at each node) of a soda straw (Roux et al. 2014). Linear dynamic differential equations are solved using Newmark's numerical integration method (Newmark 1959). The geometrical and mechanical parameters of the beam are given in Sect.~\ref{sec:dyn}. We neglected water drops hanging at the soda straw extremity---when present---as they weigh generally less than 1--2\% of the speleothem own weight. During the transient response analysis, the solicitation consists of the ground velocity signal \GPC{} derived from Shot A recorded at sensor GPa (see Sect.~\ref{sec:seismic} and Fig.~\ref{fig:signal}a). This ground motion is applied transversally at the speleothem embedded extremity. Preliminary tests were performed to ensure the numerical convergence of the solver so that a $10^{-5}$~s time step and 20 finite elements per soda straw were selected. At each time step, the motion and stress state at each node of the soda straw was recorded.

\begin{figure}[htbp]
\centering
\includegraphics[width=\textwidth]{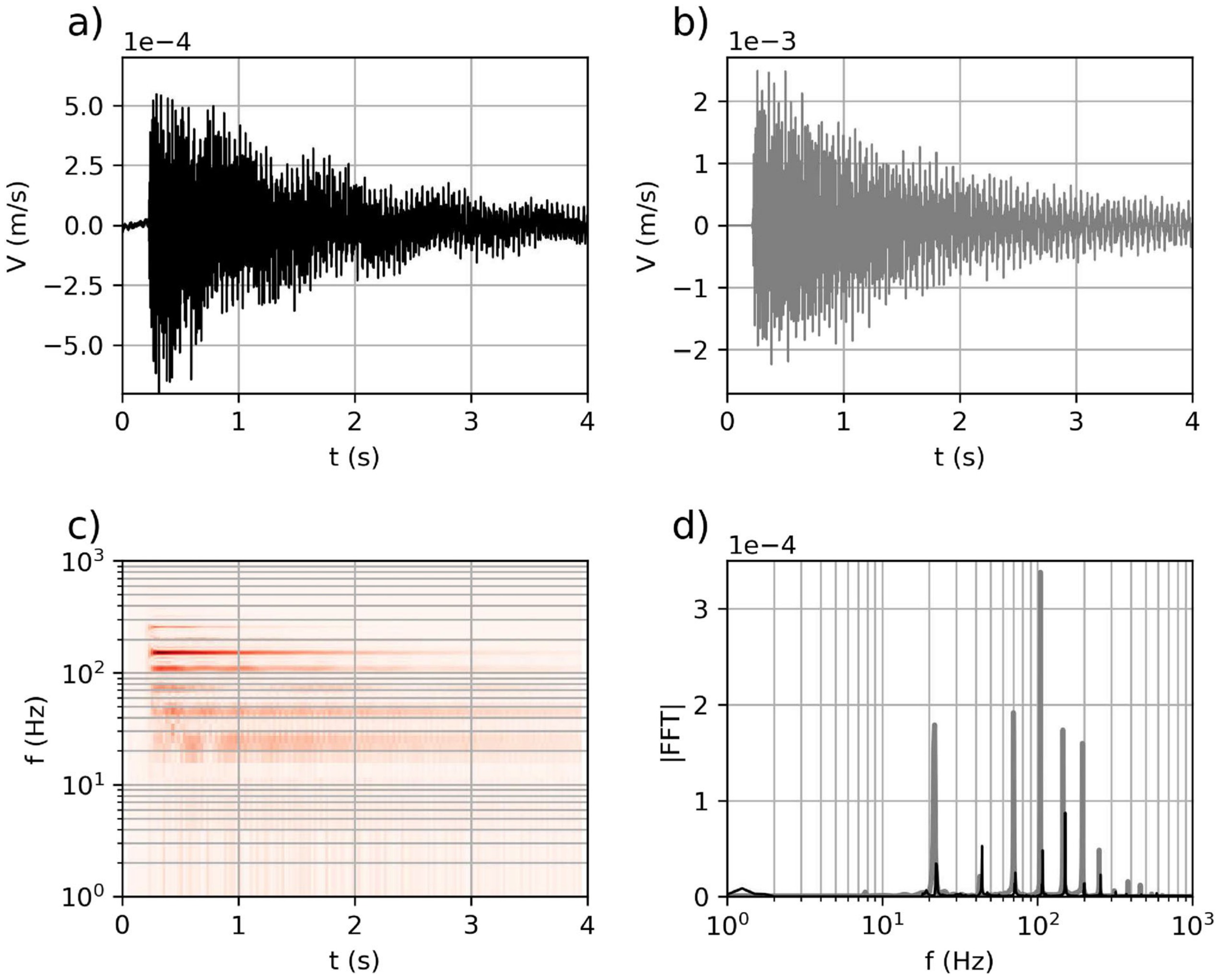}
\caption{Motion of CHO\_D3.01 soda straw tip (length $L$ = 1.78~m). a) In situ laser vibrometer measurements during explosive shot A. b) Numerical FEM simulation results with signal \GPC{} as input. c) Spectrogram signal shown in a. d) Fourier spectra (black: laser measurements; gray: FEM modeling).}
\label{fig:tip}
\end{figure}

We first simulated the CHO\_D3.01 ($L$ = 1.78~m) soda straw response during \GPC{} solicitation. The soda straw free extremity motion is shown in Fig.~\ref{fig:tip} for both experimental laser measurements (black) and FEM transient analysis (gray). Time signals are shown on the top and corresponding spectra at the bottom. Both time signals show a sharp increase with peak amplitude followed by a smooth decay. The signal duration is consistent between both signals and lasts a few seconds. The amplitude of the numerical tip motion ($\sim$2~mm/s) yet appears about 4 times greater than the experimental motion (about 0.5~mm/s). This overestimation could originate from multiple factors. Firstly, the soda straw tip motion may not be aligned along the laser line of sight as the laser position was constrained by the lake and the speleothem tip motion direction was not known a priori. We estimate that the angular bias could cause up to 25\% motion underestimation in the experimental laser records. Secondly, the solicitation \GPC{} applied to the numerical soda straw was estimated from records made along the cave wall (sensor GPa, Fig.~\ref{fig:site}), which is located about 25~m closer to the shot than the soda straw CHO\_D3.01. Based on the propagation law later shown in Sect.~\ref{sec:charge}, this distance effect may represent a factor of $\sim$2 in amplitude. In the rest of this study, we hence retained this factor 4 as a safety factor.

The spectral content for both experimental (black) and numerical (gray) tip motions are shown in Fig.~\ref{fig:tip}d. We observe a very good agreement between experimental and numerical natural frequencies ($f_0$--$f_{13}$, see also Table~\ref{tab:dyn}), especially in the range 10--300~Hz, which is most excited by the blast (Fig.~\ref{fig:signal}). These resonance frequencies values also fit almost exactly the Euler-Bernoulli beam equations. The narrowness of the measured spectral peaks reflects a low damping ratio at the natural frequencies, which is consistent with the laboratory characterization (Sect.~\ref{sec:dyn}) and satisfactorily simulated by FEM. These results suggest the appropriateness of the FEM dynamic model to simulate the response of CHO\_D3.01 soda straw, as its spectral content, damping, and transient response are correctly reproduced.

We then modeled the Choranche soda straw population in the lake room with speleothem length $L$ varying between 0.025 and 3.0~m with a step of 0.025~m. For each simulation, each computation time step and each soda straw length, the maximal tensile stress $\sutmax$ in the speleothem was saved. $\sutmax$ value takes into account both the static stress due to speleothem own weight and the dynamic stress due to shaking. This study is the first to simulate resonance amplification when applicable, unlike previous work (see Table~\ref{tab:lit}).

\begin{figure}[htbp]
\centering
\includegraphics[width=0.75\textwidth]{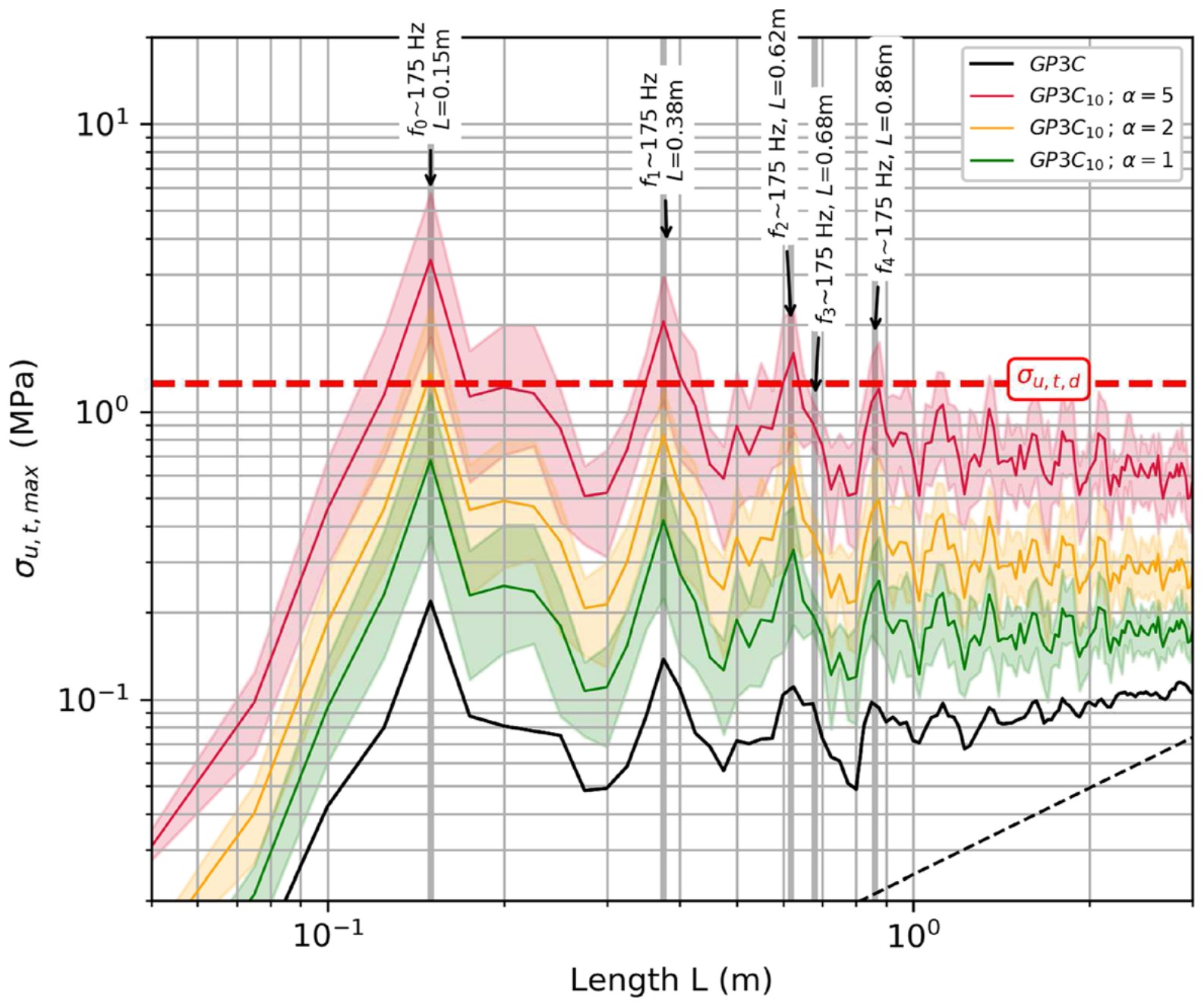}
\caption{Maximal tensile stress $\sutmax$ as a function of speleothem length computed in FEM numerical simulations. Black: signal \GPC{} as input. Colored areas: signal \GPCten{} as input, for several multiplication factors $\alpha$. Mean value: continuous line, min/max values: filled area. Vertical gray lines and black arrows: natural frequencies associated with stress peaks. Tensile resistance design value $\sutd = 1.25$~MPa derived from laboratory measurements is shown as horizontal dashed red line. Tensile stress due to speleothem own weight is shown as black dashed line (bottom right corner).}
\label{fig:stress}
\end{figure}

Figure~\ref{fig:stress} shows the maximal tensile stress $\sutmax$ as a function of speleothem length $L$. The soda straw own weight is shown as dashed black line. We observe that the stress level remains limited during test shot \GPC{} (black continuous line), peaking at about 0.2~MPa for a 0.15~m long speleothem. For soda straws in the range 0.025--1~m, we observe successive stress peaks at specific lengths (black arrows and gray vertical lines). These spectral peaks correspond to soda straws natural frequencies ($f_0$--$f_4$), which are very close to the maximal spectral content of signal \GPC{} ($\sim$150--200~Hz, with a peak at about 175~Hz, see Fig.~\ref{fig:signal}c). For longer soda straws, the maximal stress appears flatter, showing only little increases with length due to the speleothem own weight. In other words, most of short soda straws ($L < 1$~m) experience greater stress than the longer ones ($L > 1$~m) under this test blast A because of dynamic amplification.

As the peak in stress may correspond to a higher mode (i.e. not the fundamental mode $f_0$), the maximal stress is not necessarily located at the speleothem embedded extremity. This observation is in contradiction with the equivalent static approach ESA (e.g. Cadorin et al. 2001), which states that the maximal stress is due to flexural bending and is located at the embedded extremity of the speleothem. In the ESA, the stress increases with the squared speleothem length ($L^2$) for a given ground acceleration (the speleothem mass that controls the inertia force amplitude scales with $L$ and the moment of this inertia force also scales with $L$).

We then carried out simulations for delayed blast shots, which are almost always used for blasting excavation works. It consists in firing several explosive charges separated by a few tens of milliseconds, the delayed charges lying inside the same borehole or in separated boreholes. The assumption is that delaying the successive wave will prevent constructive interaction between the wavefields (i.e. equal chances of constructive and destructive interactions). For large delays between explosive charges and stiff media (i.e. high frequency content, short signal duration), a complete decoupling between charges may even be achieved (Chapot 1981a). The quantification of the optimum delay yet remains delicate (see Hoshino et al. 2000) and standard 25~ms delays are commonly used for civil engineering excavation works.

We created a synthetic signal \GPCten, which simulates 10 explosive charges of 0.2~kg MIC (maximal instantaneous charge of explosive). Each charge was separated by a delay randomly ranging between 20 and 30~ms, to take into account variations in detonator accuracy and propagation times. Ten simulations were carried out to estimate the effect of random delay wandering. An example of \GPCten{} is shown in Fig.~\ref{fig:signal}b (time signal) and in Fig.~\ref{fig:signal}c (spectrum, gray curve). We observe that \GPCten{} signal culminates at about 1.3~mm/s, which is comparable with \GPC{} peak ground velocity. \GPCten{} yet shows a spectrum richer in high frequencies compared with single blast \GPC.

The maximal tensile stress $\sutmax$ during \GPCten{} (Fig.~\ref{fig:stress}, $\alpha = 1$) wanders across the green area during the 10 random simulations. Stress average value is shown by the continuous green line. We observe that the peak stress increases of a factor about 3.5 for \GPCten{} (green line) compared with \GPC{} (black line, Fig.~\ref{fig:stress}), although the signal peak velocity increased only slightly. This effect is probably due to the successive seismic wavefronts, which tend to enrich the signal content in high frequencies (Fig.~\ref{fig:signal}c), hence increasing the ground acceleration level. Such negative impact of delayed charges on speleothem integrity was also pointed out by Lacave et al. (2003), who advised delays as large as 500~ms for underground blasts to avoid superposition effects. This delay is probably greater for soda straws, which have very low internal damping and slender shape. Unfortunately, electrical detonators used for this study were available only with 25-ms time steps.

Varying the input signal amplitude by a coefficient $\alpha$, we observed that the stress level appeared linearly related to the input signal amplitude, for a given signal (i.e. for a specific frequency content). Figure~\ref{fig:stress} shows the results for \GPCten{} input signal and amplitude factors $\alpha = 1$, $\alpha = 2$, and $\alpha = 5$. The tensile failure design value $\sutd = 1.25$~MPa (see Sect.~\ref{sec:rupture}) is reached for $\alpha = 2$, which means a \GPCten{} PPV$_{\max}$ of about 2.4~mm/s. The peak in tensile stress occurs for $\sim$0.15~m long soda straws, which experience dynamic amplification at $f_0$.

In order to test the influence of different input signals, we conducted these simulations for other test shots B and C, which have different MIC and location (see Fig.~\ref{fig:site}c). The same processing steps as shot A were applied to derive corresponding \GPC{} and \GPCten{} signals. Maximal tensile stress $\sutmax$ as a function of ground peak particle velocity (PPV) is shown in Fig.~\ref{fig:ppv}, for both single charge case (black dots, input signals \GPC) and delayed charge case (blue dots, input signals \GPCten). Results for blast test A (see Fig.~\ref{fig:stress}) are shown with squares in Fig.~\ref{fig:ppv}. A least square linear fit for all blast tests is drawn as a dashed line. Whatever the blast shot and its frequency content, the maximal tensile stress $\sutmax$ increases with ground velocity for both single shots (Fig.~\ref{fig:ppv}a) and delayed charges blasts (Fig.~\ref{fig:ppv}c). We yet note that our study lacks from too few test shots and especially from strong motion blast tests, for cost and safety reasons. We estimated approximately the maximal admissible peak ground velocity PPV$_{\max}$ as the abscissa where the regression (dashed lines) intercepts the design resistance $\sutd = 1.25$~MPa (dashed red line). PPV$_{\max}$ is hence estimated at about 8.6~mm/s for single charge blasts and about 2.4~mm/s for delayed shots. These values are considered along the cave wall, i.e. at the soda straw embedded extremity.

\section{Estimation of maximal charge}
\label{sec:charge}

In order to estimate the maximal explosive charge that can be fired for the excavation, we fitted a propagation law on the vibration peak particle velocities (PPV). Measurements from three component geophones, GPa, GPb, and GPc, for all test shots were used in combination with an additional seismic profile (PS) located outside the cave (Fig.~\ref{fig:site}c). PS profile was composed of ten 4.5~Hz vertical geophones cemented on the rock with plaster.

\begin{figure}[htbp]
\centering
\includegraphics[width=\textwidth]{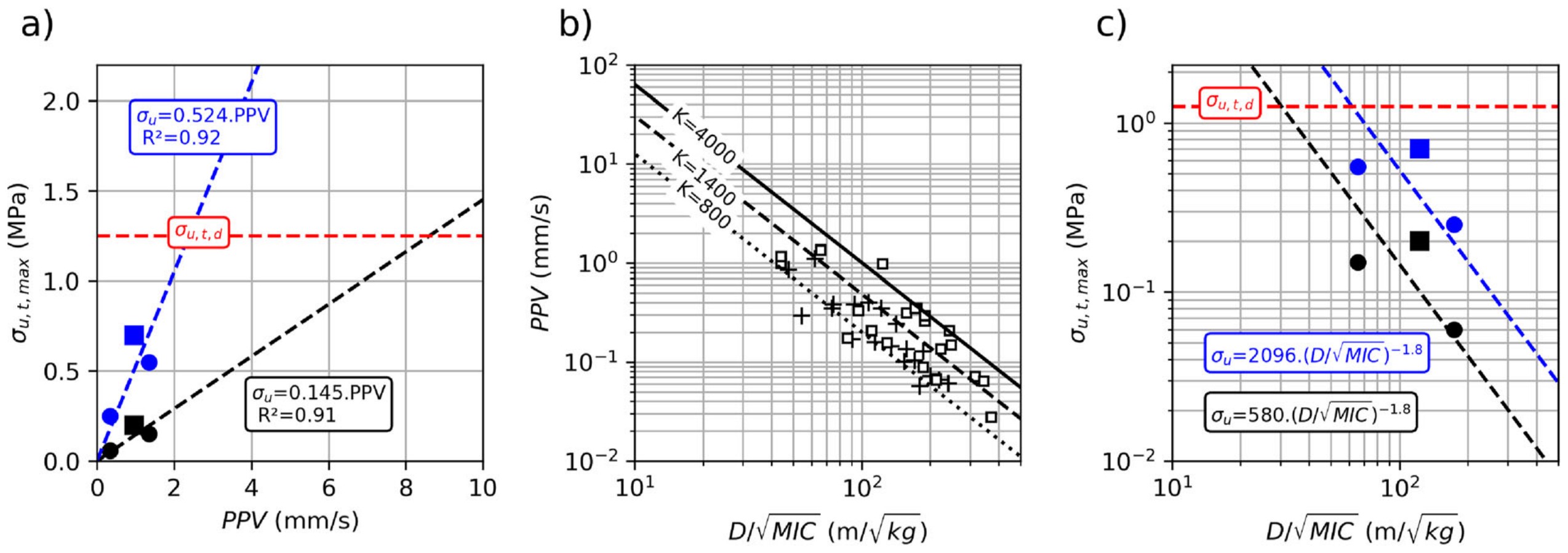}
\caption{a) Maximal tensile stress $\sutmax$ computed by FEM as function of ground peak particle velocity (PPV). Single blast shots are shown in black and delayed blasts are shown in blue. \GPC{} and \GPCten{} for shot A presented in Fig.~\ref{fig:signal} are shown with squares. Least-square regressions are shown with dashed lines. b) Ground peak particle velocity (PPV) as a function of scaled distance (i.e. distance $D$ to the blast divided by square root of the maximal instantaneous explosive charge MIC). c) Maximal tensile stress $\sutmax$ computed by FEM as function of scaled distance. Single blast shots are shown in black and delayed blasts are shown in blue. \GPC{} and \GPCten{} for shot A presented in Fig.~\ref{fig:signal} are shown with squares. Least-square regressions are shown with dashed lines. For a and c, the maximal tensile stress admissible in the soda straws $\sutd$ is shown as dashed red line.}
\label{fig:ppv}
\end{figure}

The measured PPV as a function of scaled distance are shown in Fig.~\ref{fig:ppv}b. GP measurements (inside the cave) are shown with white squares, PS profile with black crosses. Peak particle velocity shows a steady decrease with the scaled distance (in log-log axes). A greater spreading is observed for GP measurements, which recorded test shots fired at different locations (Fig.~\ref{fig:site}c) than for profile PS.
\begin{equation}
PPV = K \left(\frac{D}{\sqrt{MIC}}\right)^{-b}
\label{eq:ppv}
\end{equation}
with:\\[2pt]
\begin{tabular}{@{}ll@{}}
$PPV$ & peak particle velocity (in mm\,s$^{-1}$) \\
$MIC$ & maximal instantaneous charge (in kg) \\
$D$ & distance (in m) \\
$K$ & site coefficient (without unit) \\
$b$ & site attenuation coefficient (without unit) \\
\end{tabular}

\medskip
The power-law shown in Eq.~\ref{eq:ppv} was then fitted on the measurements (Chapot 1981b; Duvall and Petkof 1958 in Kumar et al. 2016). This formula scales the distance by the square root of explosive maximal instantaneous charge (i.e. $D/\sqrt{MIC}$) to account for the mix of body and surface waves emitted. The exponent $b$ controls the attenuation and ranges generally between 1 and 2 (Kumar et al. 2016). We chose $b = 1.8$ as best fit on the experimental data. The coefficient $K$ depends both on the site and shot parameters; it tunes the PPV level for unit scaled distance. It ranges commonly between 300 and 6000, with a median value of 2500 (Chapot 1981b). In our case, $K$ ranges between 800 and about 4000 (respectively dotted and continuous lines in Fig.~\ref{fig:ppv}b). The PS profile located on the external, fractured rock skin shows the lower $K$, in agreement with the results from the GPR campaign (Sect.~\ref{sec:site}). Higher $K$ is observed for most GP measurements located inside the cave, reflecting the greater rock quality at depth. Therefore, we selected $K = 4000$ as a reasonably safe design value for the estimation of the maximal MIC admissible during the works.

Maximal tensile stress $\sutmax$ as a function of scaled distance is shown in Fig.~\ref{fig:ppv}c for both single shots (black dots) and delayed charges blasts (blue dots). Both series show a clear and consistent decrease in $\sutmax$ experienced by the speleothems when increasing $D$ and/or reducing MIC. We modified Eq.~\ref{eq:ppv} into Eq.~\ref{eq:smax} to express the maximal tensile stress $\sutmax$ as a function of scaled distance ($D/\sqrt{MIC}$). The site coefficient $K_1$ in Eq.~\ref{eq:smax} is equal to regression slope coefficients from Fig.~\ref{fig:ppv}a multiplied by $K = 4000$. $K_1$ is equal to 580 for single shots and about 2096 for delayed shots.
\begin{equation}
\sutmax = K_1 \left(\frac{D}{\sqrt{MIC}}\right)^{-b}
\label{eq:smax}
\end{equation}
with:\\[2pt]
\begin{tabular}{@{}ll@{}}
$\sutmax$ & maximal tensile stress (in MPa) \\
$MIC$ & maximal instantaneous charge (in kg) \\
$D$ & distance (in m) \\
$K_1$ & site coefficient (without unit) \\
$b$ & site attenuation coefficient (without unit) \\
\end{tabular}

\medskip
The critical scaled distance ($D/\sqrt{MIC}$) is reached when the power law equals the design resistance of the speleothem $\sutd = 1.25$~MPa (dashed red line). Critical scaled distances range between 23.7 (single shot) and 48.5~m/$\sqrt{\mathrm{kg}}$ (blast with delayed charges). Lacave et al. (2003) advised similar minimal scaled distances (i.e. between 50 and 70~m/$\sqrt{\mathrm{kg}}$) when using delayed explosive charges close to speleothems.

\section{Discussion}

Our study dealt only with perfectly tubular soda straws, which is the case for most of the speleothems found in the lake room. However irregular sections or local wall thickening can be found on some samples in the Choranche cave. In addition, other speleothems are present in the cave such as massive stalactites, stalagmites, columns, flowstone, and drapery. For such speleothems, the dynamic parameters may differ significantly from our study. In addition, the FEM should be performed individually and consider the actual speleothem geometry. This represents a time-consuming task, applicable only for a limited number of speleothems.

During our vibration study, we fired only underloaded shots for safety reasons so that no rock was fractured. Such blasts tend to generate higher vibration amplitudes than excavation blasts for the same explosive charge. This assumption provides an additional safety margin. In addition, test blasts generally consist in isolated shots, while delayed charges are fired for the production phase. Synthetic signals created by repeating pulses e.g. every 25~ms tend to underestimate the acceleration level applied at the speleothem embedded extremity (i.e. soda straw top in our case). We hence applied a random delay in the range 20--30~ms to consider the accuracy of electric detonators (about a few milliseconds, Silva et al. 2018) and variations in source-speleothem distance due to blast hole configuration. The nonperiodicity of the input signal created stress peaks in the soda straws. This effect appears clearly in low-damped structures such as speleothems (and more especially soda straws) for which only little energy decay occurs between successive wave train arrivals (Lacave et al. 2003).

\begin{figure}[htbp]
\centering
\includegraphics[width=\textwidth]{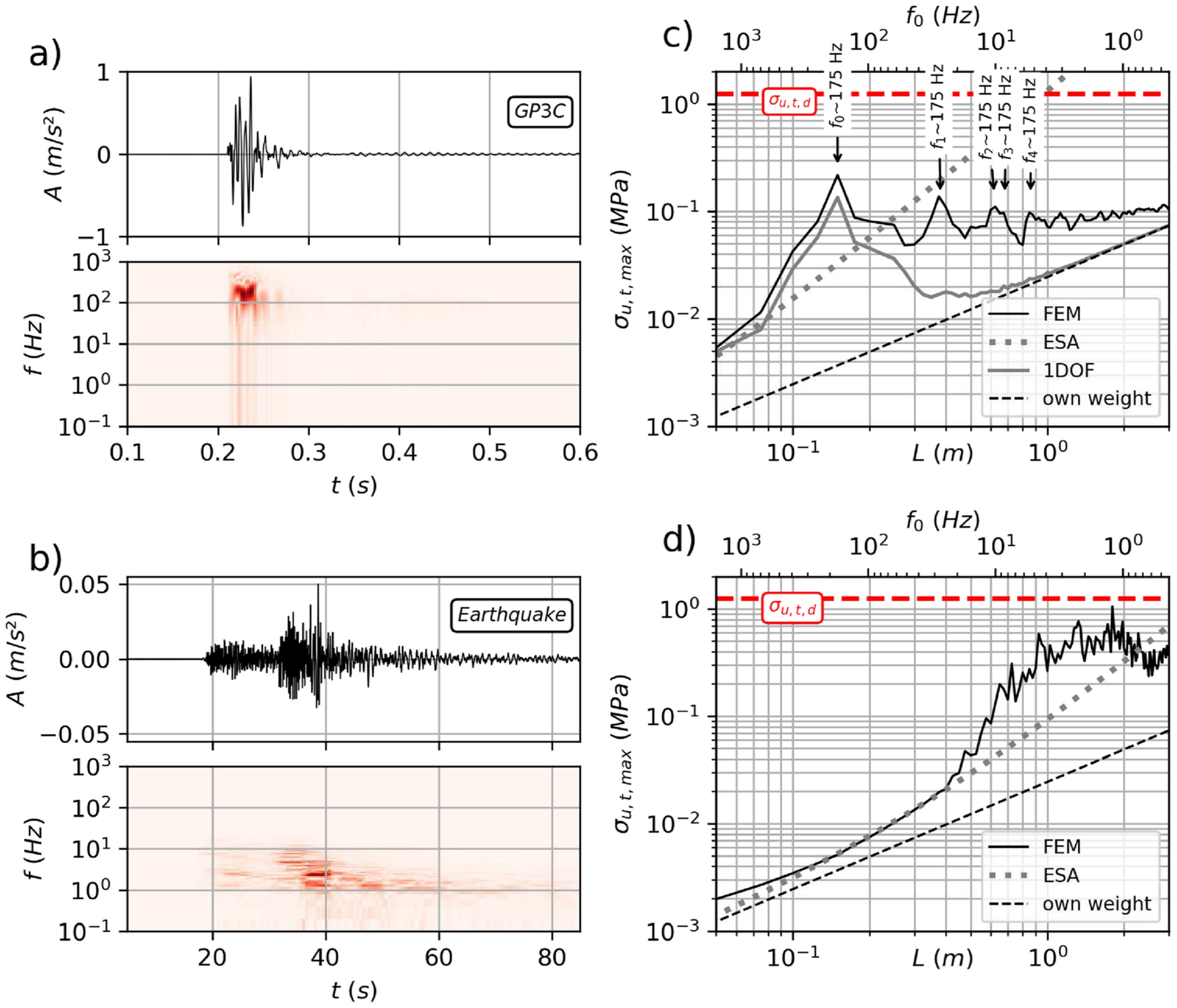}
\caption{a) Ground acceleration (top) and spectrogram (bottom) of signal \GPC{} applied at the speleothem embedded extremity. b) Ground acceleration (top) and spectrogram (bottom) of earthquake signal applied at the same location. Spectrograms are normalized between 0 (white color) and 1 (red). Subplots c and d show the maximal tensile stress $\sutmax$ as a function of speleothem length, for blast test \GPC{} (a) and earthquake (b) as input signals, respectively. Results are drawn for the dynamic FEM (continuous black line), one degree of freedom 1DOF (continuous gray line), and equivalent static approach (ESA) (thick gray dotted line). Tensile resistance design value $\sutd = 1.25$~MPa derived from laboratory measurements is shown as dashed red line. Tensile stress due to speleothem own weight is shown as dashed black line. The top x-axis shows the theoretical fundamental frequency $f_0$ for a speleothem of length $L$ (see Eq.~\ref{eq:fn} and Fig.~\ref{fig:lab}b).}
\label{fig:compare}
\end{figure}

In our study, the finite element model allowed us to simulate the dynamic amplification experienced by soda straws, whatever the excited mode. Signal \GPC{} and corresponding spectrogram recorded during single blast test A are recalled in Fig.~\ref{fig:compare}a. The maximal tensile stress in the speleothems is shown in Fig.~\ref{fig:compare}c (FEM, black curve; same curve as in Fig.~\ref{fig:stress}). Our study showed that relatively short soda straws (0.1--1~m in length) were the most exposed to nearby rock blasting because they experience higher dynamic amplification. Indeed, the conjunction between peaks in ground motion frequency content (170--180~Hz for blast A, Fig.~\ref{fig:signal}c) and soda straws resonance frequencies gives rise to significant amplification. Higher modes (up to $f_4$ for soda straw length $L$ = 0.9~m) experienced significant amplification under nearby blasting.

For comparison purposes, we computed the maximal tensile stress $\sutmax$ yielded by two other approaches during the same \GPC{} solicitation. In the first approach, we used a simple 1 degree of freedom (1DOF) spring-mass system to compute the maximal relative mass displacement. The maximal tensile stress at the embedded extremity was estimated using static Euler-Bernoulli beam (EBB) equations for embedded-free flexural test. For speleothems shorter than $\sim$0.3~m in length, the maximal tensile stress predicted by 1DOF (gray curve, Fig.~\ref{fig:compare}c) is slightly underestimated but shows consistent shape compared to FEM (black curve). Indeed, the fundamental natural frequency $f_0$ coincides with the predominant ground motion excitation ($\sim$175~Hz for shot A) for short speleothems ($<$0.3~m). In this case, the maximal stress can be approximately estimated using 1DOF dynamic model and EBB flexural equations. In contrast, the stress level yielded by 1DOF is significantly underestimated (up to a factor 5) for longer speleothems. They indeed exhibit significant dynamic motion at higher modes ($f_1$--$f_4$, see FEM results in Fig.~\ref{fig:compare}c). Those modes are not simulated by the 1DOF spring-mass model, which yields negligible mass motion (``inertial'' behavior) for speleothems longer than $\sim$0.6~m (i.e. for $f_0 < 10$~Hz, Fig.~\ref{fig:compare}c).

In the second approach, we used the equivalent static approach (ESA, gray dotted line in Fig.~\ref{fig:compare}c). The maximal tensile stress $\sutmax$ is correctly estimated for very short soda straws ($L < 0.08$~m). The rigid-body assumption is fulfilled for such speleothems as their resonance frequencies are considerably higher than the blasting spectral content (see $f_0$ in Fig.~\ref{fig:compare}c, top axis). The tensile stress is yet considerably underestimated for 0.08--0.2~m long soda straws as ESA does not take into account the dynamic amplification that occurs mainly at the fundamental mode for those speleothems ($f_0$ = 175~Hz for $L \sim 0.15$~m). In contrast, $\sutmax$ is considerably overestimated by the ESA when $L > 0.2$~m. Indeed, $\sutmax$ scales with $L^2$ in ESA as both the inertia force and the bending moment scale individually with $L$. In contrast, FEM shows lower $\sutmax$ because the rigid body assumption is not valid anymore and tends to overestimate the stress level.

We also compared the FEM, 1DOF, and ESA methods for a regional earthquake solicitation (Fig.~\ref{fig:compare}b and d). The historical reference earthquake in the Vercors massif is the 25/04/1962 Corrençon $M_L$ = 5.3--5.5, located at about 11~km in distance from the Choranche cave. Unfortunately, its seismic signal was not recorded at that time. We hence selected the 07/04/2014 Barcelonnette earthquake as a proxy ($M_L$ = 5.2, $M_W$ = 4.8, depth about 13~km/surface). The ground motion was recorded at station SURF located at $\sim$10.0~km in distance from the epicenter (RAP-RESIF network, Sira et al. 2014). The acceleration and velocity amplitudes reached peak values PGA = 0.5~m/s$^2$ and PPV = 0.044~m/s. Such event is quite rare for mainland France (once or twice per decade) but remains weak compared with possible earthquakes in the Choranche cave area (e.g. the construction design PGA for new buildings equals 1.6~m/s$^2$ in the Vercors, for comparison).

As earthquake-induced motion is strongly attenuated in underground cavities (Becker et al. 2006; Gribovszki et al. 2017), the earthquake signal amplitude was corrected following Hu and Xie (2005) formula. Site conditions for seismic station SURF belong to the soil over rock class (Hollender et al. 2015), with a mean shear wave velocity over the first 30~m of subsurface $V_{s30}$ = 420~m/s. In this case, the expected PGA in a 350-m-deep cave scales at about 1/10th the surface PGA, for both vertical and horizontal components (Hu and Xie 2005). The modified acceleration signal (top) and spectrogram (bottom) are shown in Fig.~\ref{fig:compare}b. The PGA culminates at about 0.05~m/s$^2$, with a signal frequency content ranging between 0.1 and $\sim$10~Hz. Results are shown in Fig.~\ref{fig:compare}d for FEM (continuous black line), 1DOF (continuous gray line), and ESA (dotted gray line). Maximal tensile stress curves yielded by FEM and 1DOF show a similar shape for most of the speleothems. 1DOF yet tends to underestimate the stress level as it takes only into account the contribution of the fundamental mode $f_0$ for dynamic amplification. The error is greater for longer speleothems because their upper modes $f_1$--$f_2$ may also lie in the seismic frequency range (i.e. below 10~Hz, see Fig.~\ref{fig:lab}b) contrary to Lacave et al. (2004) assumption. The ESA performs approximately well for soda straws shorter than 0.6~m, whose resonance frequencies are greater than the seismic bandwidth ($f_0 > 10$~Hz, Fig.~\ref{fig:lab}b). For such soda straws, the rigid body assumption is valid. The ESA yet underestimates $\sutmax$ when the dynamic resonance becomes non negligible ($0.6 < L < 2.5$~m) and overestimates the tensile stress for speleothems longer than 2.5~m.

For a 0.05~m/s$^2$ PGA at soda straw embedded extremity, the maximal tensile stress $\sutmax$ peaks at about 1.0~MPa (Fig.~\ref{fig:compare}d). Highest tensile stresses ($0.5 < \sutmax < 1$~MPa) are observed in speleothems measuring between 1 and 2~m in length, which are poorly simulated by 1DOF or ESA. The maximal induced stress $\sutmax$ during the earthquake remains smaller than the design value $\sutd = 1.25$~MPa in the FEM simulations so that soda straw rupture is unlikely for the selected earthquake event. Our results suggest that calcite strength design limit $\sutd$ is reached when PGA at soda straw embedded extremity exceeds about 0.06~m/s$^2$. Such PGA value may be reached for closer earthquakes, higher source magnitude, and/or lower PGA attenuation with depth, depending on the site conditions.

Our study underlines the need for a proper dynamic amplification modeling when speleothem natural frequencies and solicitation spectrum coincide, even partially. The PGA value itself does not contain enough information as the solicitation frequency content plays a major role. Our results yielded admissible PGA of about 0.06~m/s$^2$ under earthquake and about 2.4~m/s$^2$ for a blast with ten delayed explosive charges (see signal \GPCten; $\alpha = 1$ in Fig.~\ref{fig:stress}). Such relatively low values result from our adequate dynamic amplification simulation, from the intrinsic vulnerability of the soda straws, and also from the safe assumptions made in terms of calcite tensile strength ($\sutd = 1.25$~MPa) and admissible rupture probability. Under seismic solicitation, Lacave et al. (2004) found slightly higher admissible PGA (0.1--0.2~m/s$^2$) for a $\sutd$ distribution peaking at about 3~MPa. The admissible PGA was computed as 5\% rupture probability during earthquake for very vulnerable, slender, speleothems. For a slender (3~m long, 0.03~m in diameter)-filled stalagmite, an admissible PGA of 0.5~m/s$^2$ was found by Gribovszki et al. (2008) assuming $\sutd = 2.61$~MPa. For the unique previous case study of speleothems exposed to blasting works, Lacave et al. (2003) defined an admissible PGA of about 1.8~m/s$^2$, at 5\% rupture probability for soda straws.

In speleoseismology, the ESA should be restricted to low frequency solicitation (e.g. earthquakes) applied to rigid, massive speleothems with high resonance frequencies. ESA application to soda straws---whose natural frequencies may be as low as a few tenths of Hz or a few Hz---may significantly underestimate or overestimate the induced dynamic stress. The 1DOF approach performs slightly better as it simulates the dynamic amplification occurring at the fundamental mode $f_0$. It yet tends to underestimate the induced tensile stress, especially when higher modes are excited. Application of 1DOF or ESA without caution may induce severe biases in speleoseismological studies such as for earthquake hazard studies.

\section{Conclusions}

This research work was conducted to preserve remarkable soda straws exposed to nearby rock blasting in Choranche cave. We first characterized the geometric, dynamic, and ultimate properties of the speleothems through laboratory and field tests. Blast tests in situ were used as input for a dynamic 2D finite element code. We showed that the soda straw dynamic response varies considerably depending on the match between ground motion frequency content and speleothem natural frequencies. Short soda straws ($L < 1$~m) experienced higher stress under blast tests than the longer ones due to dynamic resonance. Opposite conclusions were made for speleothems exposed to a regional earthquake because of lower ground motion frequency content. This underlines the good performance of our dynamic finite element model, whatever the solicitation and excited modes. We show that the omission of dynamic resonance or its simplification as made in previous studies may significantly underestimate the induced load in speleothems.

We defined an admissible vibration threshold along the cave wall and we monitored its observance during the excavation blasts. The works were carried out without causing any harm to the soda straws. Our research work may constitute a milestone for future speleoseismological studies. As urban growth, renewal, and densification processing are increasing in many countries, we expect more and more case studies of concretionary caves exposed to anthropogenic vibrations.

\section*{Acknowledgments}

This work would not have been achieved successfully without the great support, the strong implication, and all the fruitful discussions conducted with Rémy Allier (CAN company) and David Cantalupi (Alpes Minages). The authors would like to thank Loïc Baril, Leslie Huaux, Gilbert Mantovani, and the whole Choranche team for their welcome and fruitful discussion, as well as providing access, support for the experiments, and information about the pedestrian tunnel excavation. Thanks to Didier Hantz, Axel Pionchon, Lambert Fabre, and Charles Ribeyre for their help during the fieldwork at any time during the day or the night. Thanks to the CAN company (esp. Rémy Allier), David Cantalupi (Alpes Minages), and Laurent Pinet (SEMD) for their help in the field and providing blasting information. Some of the seismic sensors were prepared and lent by GPRoge (Isabelle Douste-Bacqué and Ildut Pontdaven). Thanks to Stéphane Mercier for building a specific geophone shelf brace to reduce the L-bracket resonance. Thanks to Philippe Roux and Ludovic Moreau for their help in the laboratory experiments. Figures were drawn using Matplotlib Python package, while seismic data was handled thanks to Obspy. RESIF-RAP is a national Research Infrastructure, recognized as such by the French Ministry of higher education and research. RESIF-RAP is managed by the RESIF Consortium, composed of 18 research institutions and universities in France. RESIF additionally supported by a public grant overseen by the French national research agency (ANR) as part of the ``Investissements d'Avenir'' program (reference: ANR-11-EQPX-0040) and the French Ministry of environment, energy and sea.

\section*{Funding information}

This work was initiated and funded by the SARL Grotte de Choranche (\url{https://www.visites-nature-vercors.com/fr/choranche}). Additional funding was provided by VOR program from University Grenoble-Alpes.

\section*{References}
\begingroup
\small
\setstretch{1.1}
\setlength{\parindent}{-1.5em}
\setlength{\leftskip}{1.5em}
\setlength{\parskip}{0.3em}

Becker A, Davenport CA, Eichenberger U, Gilli E, Jeannin P-Y, Lacave C (2006) Speleoseismology: a critical perspective. J Seismol 10:371--388

Bednárik M (2009) Seismometric portrayal of calcite tubular stalactites. PhD Theses. Geophysical Institute, Slovak Academy of Sciences

Bokelmann G, Gribovszki K (2015) Constraints on long-term seismic hazard from vulnerable stalagmites. Eucentre Foundation, Pavia, p 7

Cadorin J-F, Jongmans D, Plumier A, Camelbeeck T, Delaby S, Quinif Y (2001) Modelling of speleothems failure in the Hotton cave (Belgium). Is the failure earthquake induced? Geol Mijnb 80:315--322

Chapot T (1981a) Etude des vibrations provoquées par les explosifs dans les massifs rocheux. RAPP RECH LPC NO 105

Chapot T (1981b) Study of vibrations caused in rock masses by explosives. Rapport de Recherche LPC

Chopra AK (2001) Dynamics of structures: theory and applications to earthquake engineering. Pearson, Hoboken, NJ. Prentice-Hall international series in civil engineering and engineering mechanics. Fifth edition

Crispim J (1999) Seismotectonic versus man-induced morphological changes in a cave on the Arrabida chain (Portugal). Geodin Acta 12:135--142

Delannoy J-J (1986) Contribution à l'étude des circulations aquifères dans le géosystème Coulmes-Choranche; Présentation du site expérimental de la Grotte de Coufin. Rev Géograph Alpine 74:83--92

Delannoy J-J, Gauchon C, Hobléa F, Jaillet S, Maire R, Perrette Y, Perroux A-S, Ployon E, Vanara N (2009) Karst: from palaeogeographic archives to environmental indicators. Géomorphologie 15:83--94

Duvall WI, Petkof B (1958) Spherical propagation of explosion-generated strain pulses in rock. Technical Report. Research organization: Bureau of Mines. Report number: BM-RI-5483

Fairchild IJ, Baker A (2012) Speleothem science: from process to past environments. Wiley-Blackwell. p 450. \url{https://doi.org/10.1002/9781444361094}

Ferranti L, Pace B, Valentini A, Montagna P, Pons-Branchu E, Tisnérat-Laborde N, Maschio L (2019) Speleoseismological constraints on ground shaking threshold and seismogenic sources in the Pollino range (Calabria, southern Italy). J Geophys Res Solid Earth 124:5192--5216. \url{https://doi.org/10.1029/2018JB017000}

Forti P (2001) Seismotectonic and paleoseismic studies from speleothems: the state of the art. Geologica Belgica 4(3--4):175--185

Forti P, Postpischl D (1980) The hypothesis of the induced activity of the faults as a result of a statistical analysis of stalagmites. In Proc Europ Reg Speleol Conf Sofia, pp. 34--39

Gauchon C, Ployon E, Delannoy J-J, Hacquard S, Hobléa F, Jaillet S, Perrette Y (2007) The concepts of heritage and heritage resource applied to karsts: protecting the Choranche caves (Vercors, France). Acta Carsol 35(2--3):2006. \url{https://doi.org/10.3986/ac.v35i2-3.226}

Gilli É (2005) Review on the use of natural cave speleothems as palaeoseismic or neotectonics indicators. Compt Rendus Geosci 337:1208--1215

Gilli E, Levret A, Sollogoub P, Delange P (1999) Research on the February 18, 1996 earthquake in the caves of Saint-Paul-de-Fenouillet area (eastern Pyrenees, France). Geodin Acta 12:143--158

Gribovszki K, Paskaleva I, Kostov K, Varga P, Nikolov G (2008) Estimating an upper limit on prehistoric peak ground acceleration using the parameters of intact speleothems in caves in southwestern Bulgaria. pp. 287--308

Gribovszki K, Bokelmann G, Szeidovitz G, Paskaleva I, Brimich L, Kovács K (2013a) Comprehensive investigation of intact, vulnerable stalagmites to estimate an upper limit on prehistoric horizontal ground acceleration. Vienna Congress on Recent Advances in Earthquake Engineering and Structural Dynamics 10

Gribovszki K, Brimich L, Varga P, Kovács K, Shen C-C, Kele S, Török Á, Novák A (2013b) Estimation of an upper limit on prehistoric horizontal peak ground acceleration using the parameters of intact stalagmites and the mechanical properties of broken stalagmites in Domica cave, Slovakia. In: Geophysical Research Abstracts, vol. 15, EGU2013-11541, 2013 EGU General Assembly

Gribovszki K, Kovács K, Mónus P, Bokelmann G, Konečný P, Lednická M, Moseley G, Spötl C, Edwards RL, Bednárik M (2017) Estimating the upper limit of prehistoric peak ground acceleration using an in situ, intact and vulnerable stalagmite from Plavecká priepast cave (Detrekői-zsomboly), Little Carpathians, Slovakia---first results. J Seismol 21:1111--1130

Hollender F, Cornou C, Dechamp A, Burnouf C, Renalier F, Thomassin S, Roullé A (2015) Caractérisation des conditions de site de 33 stations du Réseau Accélérométrique Permanent français (RAP) / Site condition characterization of 33 stations from the French permanent Accelerometric network (RAP). Marne-la-Vallée

Hoshino T, Mogi G, Shaoquan K (2000) Optimum delay interval design in delay blasting. Fragblast 4:139--148

Hu J, Xie L (2005) Variation of earthquake ground motion with depth. Acta Seismol Sin 18:72--81

Inman DJ (2008) Engineering vibration. Pearson Education Inc, Upper Saddle River, NJ

Kagan EJ et al (2017) Broken speleothems reveal Holocene and Late Pleistocene paleoearthquakes in Northern Calabria, Italy. Quaternary International. \url{https://doi.org/10.1016/j.quaint.2016.10.023}

Konečný P, Lednická M, Souček K, Staš L, Kubina L, Gribovszki K (2015) Determination of dynamic Young's modulus of vulnerable speleothems. Acta Montan Slovaca 20:156--163

Kumar R, Choudhury D, Bhargava K (2016) Determination of blast-induced ground vibration equations for rocks using mechanical and geological properties. J Rock Mech Geotech Eng 8:341--349

Lacave C, Levret A, Koller M (2000) Measurements of natural frequencies and damping of speleothems. In Proc. of the 12th world conference on earthquake engineering, Auckland, New Zealand

Lacave C, Koller M, Eichenberger U, Jeannin P-Y (2003) Prevention of speleothem rupture during nearby construction. Environ Geol 43:892--900

Lacave C, Koller MG, Egozcue JJ (2004) What can be concluded about seismic history from broken and unbroken speleothems? J Earthq Eng 8:431--455

Méjean P, Garduño-Monroy V-H, Pinti DL, Ghaleb B, Bouvier L, Gomez-Vasconcelos MG, Tremblay A (2015) U--Th dating of broken speleothems from Cacahuamilpa cave, Mexico: are they recording past seismic events? J S Am Earth Sci 57:23--31

Mogi G, Hoshino T, Kou S-Q (2000) Reduction of blast vibration by means of sequentially optimized delay blasting. In: Explosives and Blasting Technique. Balkema, Rotterdam

Newmark NM (1959) A method of computation for structural dynamics. J Eng Mech Div 85(3):67--94

Perrette Y, Jaillet S (2010) Spatial distribution of soda straws growth rates of the Coufin cave (Vercors, France). Int J Speleol 39:61--70

Postpischl D, Agostini S, Forti P, Quinif Y (1991) Palaeoseismicity from karst sediments: the ``Grotta del Cervo'' cave case study (Central Italy). Tectonophysics 193:33--44

Reynard E, Hobléa F, Cayla N, Gauchon C (2011) Iconic sites for alpine geology and geomorphology. Rediscovering heritage? J Alpine Res 99--2. \url{https://doi.org/10.4000/rga.1412}

Roux P, Guéguen P, Baillet L, Hamze A (2014) Structural-change localization and monitoring through a perturbation-based inverse problem. J Acoust Soc Am 136:2586--2597

Rustan AP (1998) Micro-sequential contour blasting---how does it influence the surrounding rock mass? Eng Geol 49:303--313

Silva J, Li L, Gernand JM (2018) Reliability analysis for mine blast performance based on delay type and firing time. Int J Min Sci Technol 28:195--204

Sira C, Schlupp A, Schaming M, Chesnais C, Cornou C, Dechamp A, Delavaud E, Maufroy E (2014) Séisme de Barcelonnette du 7 avril 2014. Rapport du BCFS 76

Szeidovitz G, Surányi G, Gribovszki K, Bus Z, Leel-Őssy S, Varga Z (2008) Estimation of an upper limit on prehistoric peak ground acceleration using the parameters of intact speleothems in Hungarian caves. J Seismol 12:21--33

Xie L, Lu W, Gu J, Wang G (2018) Excavation method of reducing blasting vibration in complicated geological conditions. Shock Vib 2518209:12. \url{https://doi.org/10.1155/2018/2518209}

\endgroup

\end{document}